\documentclass[12pt,preprint]{aastex}

\def\bge{\begin{equation}}
\def\ene{\end{equation}}
\def\bg{\begin{eqnarray}}
\def\en{\end{eqnarray}}
\shorttitle{EoS for massive neutron stars}
\shortauthors{Katayama, Miyatsu, Saito}

\begin{document}


\title{Equation of state for massive neutron stars}


\author{Tetsuya Katayama, Tsuyoshi Miyatsu, and Koichi Saito}
\affil{Department of Physics, Faculty of Science and Technology \\
Tokyo University of Science (TUS), Noda 278-8510, Japan}

\email{koichi.saito@rs.tus.ac.jp}






\begin{abstract}
Using relativistic Hartree-Fock approximation, we investigate the properties of the neutron-star matter in detail. In the present calculation, 
we consider not only the tensor coupling of 
vector mesons to octet baryons and the form factors at interaction vertexes but also the internal (quark) structure change of baryons in dense matter.  
The relativistic Hartree-Fock calculations are performed in two ways: one is the calculation with the coupling constants determined by SU(6) (quark model) symmetry, the other is 
with the coupling constants based on SU(3) (flavor) symmetry.  For the latter case, we use the latest Nijmegen (ESC08) model. 
Then, it is very remarkable that the particle composition of the core matter in SU(3) symmetry is completely different from that in SU(6) symmetry.  
In SU(6) symmetry, all octet baryons appear in the density region below $\sim 1.2$ fm$^{-3}$, while, in the ESC08 model, only the $\Xi^-$ hyperon is produced. 
Furthermore, the medium modification of the internal baryon structure hardens the equation of state for the core matter.  
Taking all these effects into account, we can obtain the maximum neutron-star mass which is consistent with 
the recently observed mass, $1.97 \pm 0.04 M_\sun$ (PSR J1614-2230).  
We therefore conclude that the extension from SU(6) symmetry to SU(3) symmetry in the meson-baryon couplings and the internal baryon-structure 
variation in matter certainly enhance the mass of neutron star. Furthermore, the effects of the form factor at vertex and 
the Fock contribution including the tensor coupling due to the vector mesons are indispensable to describe the core matter.  
In particular, the Fock term is very vital in reproducing the preferable value of symmetry energy, $a_4 \, (\simeq 30 - 40$ MeV), in nuclear matter. 
\end{abstract}





\section{Introduction}

Neutron stars are composed of the densest form of hadrons (baryons and mesons) and 
leptons.  However, the detail of particle fractions and properties of the core of  
neutron star are not yet theoretically understood.  Observations of the mass and/or radius of 
neutron star can provide the stringent constraint on the equation of state (EoS) of 
dense matter in the core.  The best determined pulsar mass is $1.4414 \pm 0.0002 M_\sun$ 
(the Hulse-Taylor pulsar), and masses of most other pulsars are found to be close to this 
canonical value.  However, several neutron stars with heavier masses have recently been discovered.  Radio timing 
observations of three Post Keplerian parameters led to the most precise measurement of the mass of 
a millisecond pulsar of  $1.667 \pm 0.021 M_\sun$ (PSR J1903-0327) \citep{freire}.  Furthermore, Shapiro delay measurements 
from radio timing observations of the binary millisecond pulsar (PSR J1614-2230) have indicated 
a mass of $1.97 \pm 0.04 M_\sun$ of the neutron star \citep{demorest}.  

Nuclear forces play a very important role in the core of neutron star, acting in concert with gravitational forces 
to form a compact object.  The observed mass of a neutron star cannot be understood without counting 
the pressure arising from the strong nuclear force, which opposes gravitational collapse.   
The EoS for the core matter has been firstly investigated in terms of only nucleons ($N$) and leptons ($\ell$).  
Such a EoS can be very stiff, and thus yield the maximum allowable mass of a neutron star such as $M_{max}  \simeq 2.8 M_\sun$, 
which is close to the absolute upper bound on $M_{max}$ stemming from causality \citep{haensel}.  
However, in dense matter, hyperons ($Y$) generated through weak interaction may also participate in the EoS. 
This is true even at low density, in the region around twice the normal nuclear matter density, $n_0 (= 0.15$fm$^{-3})$, because some models predict the low threshold density 
for the creation of hyperons in beta equilibrium.  Therefore, there is no reason to deny the appearance of hyperons in dense matter.  
When hyperons are considered, the inevitable softening 
of the EoS occurs, which implies that the maximum neutron-star mass becomes $M_{max}  \lesssim 1.5 M_\sun$ \citep{burgio,vidana}.  Such a low maximum mass is only 
marginally consistent with the  Hulse-Taylor pulsar, but is already contradicted by the pulsars of PSR J1903-0327 and PSR J1614-2230.  

To explain the EoS for a neutron star, many people have recently used Quantum Hadrodynamics (QHD) \citep{serot1984} in relativistic mean-field (RMF) or 
relativistic Hartree (RH) approximation, in which the baryons are treated as point-like objects and 
several non-linear, self-interaction terms of the scalar ($\sigma$) and/or vector ($\omega, \vec{\rho}, \phi$) mesons are 
included to push the threshold of the hyperon appearance to higher densities and obtain a heavy neutron-star mass \citep{todd,schramm,shen,bednarek,weissenborn2012a}.  

In general, an effective field theory at low energy will contain an infinite number of interaction terms, which incorporate the 
{\it compositeness} of hadrons, and it is then expected to involve numerous couplings which may be nonrenormalizable.  
{\it Naive-dimensional analysis} (NDA) \citep{manohar} can provide an organizing principle to make sensible calculations for such system.  For the nuclear many-body problem, 
the power counting scheme based on NDA \citep{furnstahl1997,furnstahl2000} allows us to expand the Lagrangian (or the total energy) in terms 
of the ratios of scalar and vector densities to the factor $f_\pi^2\Lambda$ (the pion decay constant, 
$f_\pi$, and the large mass scale, $\Lambda \sim 1$GeV).  Such ratios are usually between 
$1/4$ and $1/7$ around $n_0$, which can certainly serve as expansion parameters.  
From the point of view of the power counting, the tensor force due to the $\pi$- and $\rho$-meson exchanges, which plays a very crucial role in 
nuclear forces, can, however, be viewed as a higher-order correction in a nuclear matter, and thus the tensor contribution to the total energy is not large in usual nuclei.  However, 
in the core of neutron star where the density can grow up to $\sim 10 n_0$, the expansion supported by the power counting is no longer valid, 
because the convergence becomes worse with increasing density. 
Thus, it is interesting to study how the tensor interaction affects the properties of the dense core matter.  To take it into account, it is necessary to 
consider the Fock term, as well as the Hartree term.  The Fock contribution also ensures the effect of anti-symmetrization for baryon wave functions in matter, which 
is not included in the RH calculations.  Therefore, it is highly desirable to calculate the EoS within relativistic Hartree-Fock (RHF) approximation. 

At sufficiently high density, it seems unlikely that hadrons will maintain their identity as confined entities of quarks and gluons, but the interaction between overlapping 
hadrons will result in partial or full deconfinement.  Even around $n_0$, it is well recognized that the value of the quark condensate, $\langle {\bar q} q \rangle$, 
is {\it not} the same as that in vacuum, and that it decreases by about $20-30\%$ at $n_0$ \citep{cohen,tsushima}.  Because the quark condensate has {\it Lorentz-scalar} character, 
the decrease in $\langle {\bar q} q \rangle$ directly modifies the quark mass 
and thus leads to the change of quark wave function in a hadron.  It implies that the hadron properties at finite density are certainly different from those in vacuum. 
Thus, even in the density region below the deconfinement of quarks and gluons, the EoS should contain all this physics. 

In the quark-meson coupling (QMC) model \citep{guichon1988,saito1994a}, the properties of nuclear matter can be self-consistently calculated by the coupling of 
scalar ($\sigma$) and vector  ($\omega$) fields to the {\it quarks} 
within the nucleons, rather than to the nucleons themselves.  As a result of the scalar coupling to the quark, the internal structure of the nucleon 
is modified with respect to the free case. 
In particular, the decrease of the quark mass caused by the attractive scalar field reduces the quark scalar density in the nucleon.  
Because the quark scalar density itself is the source of the $\sigma$ field, this provides a mechanism for the saturation of 
nuclear matter, where the quark structure plays a vital role.  

In the past few decades, the QMC model has been extensively developed and applied to various nuclear phenomena with tremendous 
success \citep{guichon1996,saito1996,saito1997,saito2007}.  
In fact, the evidence for the medium modification 
of nucleon structure in a nucleus was observed in polarization transfer measurements in the quasi-elastic $(e, e^\prime p)$ reaction at the Thomas Jefferson Laboratory, 
and the result supports the prediction of the QMC model \citep{brooks}. 
It also seems vital to consider the internal structure change of the nucleon to understand the European Muon Collaboration (EMC) effect \citep{geesaman,saito1994b,cloet}.  

The QMC model has recently been extended to include the quark-quark hyperfine interactions due to the exchanges of gluon and pion based on 
chiral symmetry \citep{nagai}, 
and it has been applied to hyperons in a nuclear medium \citep{miyatsu2009}.  We call this new version the chiral QMC (CQMC) model.  
The hyperfine interaction plays an important role in the in-medium baryon 
spectra.  In particular, it can explain why the $\Lambda$ feels more attractive force than the $\Sigma$ or $\Xi$ in matter.  
The QMC model can offer a unique opportunity to self-consistently investigate the composition of the core of neutron star in terms of the quark and gluon degrees of freedom.  

In the previous calculation by \citet{miyatsu2012}, using the QHD, QMC and CQMC models, 
we have briefly reported the effects of Fock term, tensor coupling of vector mesons and baryon structure 
variation on the EoS for the core matter \citep{stone2007,whitten}.  
In the present paper, we fully investigate how the Fock term including the tensor interaction and the baryon structure change contribute to the 
EoS in RHF approximation.  We also study the effect of form factors at interaction vertexes.  Furthermore, using the latest Nijmegen (ESC08) model \citep{esc08}, 
we examine the extension from SU(6) (quark model) symmetry to SU(3) (flavor) symmetry in the meson-baryon coupling constants.  This extension leads to the significant 
changes in the particle fractions in the core matter and thus in the maximum mass. Our findings are summarized in the last section. 

\section{Relativistic Hartree-Fock calculation}

We present the relativistic formulations of QHD and QMC models in Hartree-Fock approximation. 

\subsection{\label{rhf} QHD with non-linear $\sigma$ terms}

In QHD, the octet baryons (proton ($p$), neutron ($n$), $\Lambda$, $\Sigma^{+0-}$, $\Xi^{0-}$) and mesons ($\sigma$, $\omega$, ${\vec \pi}$, ${\vec \rho\,}$) are 
assumed to be {\it structureless}.  Then, the Lagrangian density is given by
\bge
	\mathcal{L} = \mathcal{L}_{B} + \mathcal{L}_{\ell} + \mathcal{L}_{M} + \mathcal{L}_{int} ,
	\label{eq:Lagrangian1}
\ene
where
\bge
	\mathcal{L}_{B} = \sum_{B \ni p, n, \Lambda, \Sigma, \Xi}\bar{\psi}_{B}(i\gamma_{\mu}\partial^{\mu}-M_{B})\psi_{B}, \ \ \ \ \ 
	\mathcal{L}_{\ell} = \sum_{\ell \ni e^{-}, \mu^{-}}\bar{\psi}_{\ell}(i\gamma_{\mu}\partial^{\mu}-m_{\ell})\psi_{\ell},
	\label{eq:Lagrangian-baryon-lepton}
\ene
with $\psi_{B (\ell)}$ the baryon (lepton) field and $M_{B} \, (m_{\ell})$ the baryon (lepton) mass in vacuum. 
The sum $B$ runs over the octet baryons, and the sum $\ell$ is for the leptons, $e^{-}$ and $\mu^{-}$.  For the free baryon masses, we take $M_{N}=939$ MeV, 
$M_{\Lambda}=1116$ MeV, $M_{\Sigma}=1193$ MeV and $M_{\Xi}=1313$ MeV, respectively.

The meson term reads
\bg
	\mathcal{L}_{M} &=& \frac{1}{2}\left(\partial_{\mu}\sigma\partial^{\mu}\sigma-m_{\sigma}^{2}\sigma^{2}\right)
	+ \frac{1}{2}m_{\omega}^{2}\omega_{\mu}\omega^{\mu} - \frac{1}{4}W_{\mu\nu}W^{\mu\nu} \nonumber \\
	&+& \frac{1}{2}m_{\rho}^{2}\vec{\rho}_{\mu}\cdot\vec{\rho}^{\,\mu} - \frac{1}{4}\vec{R}_{\mu\nu}\cdot\vec{R}^{\mu\nu}
	+ \frac{1}{2}\left(\partial_{\mu}\vec{\pi}\cdot\partial^{\mu}\vec{\pi}-m_{\pi}^{2}\vec{\pi}^{2}\right),
	\label{eq:Lagrangian-meson}
\en
with
\bge
	W_{\mu\nu} = \partial_{\mu}\omega_{\nu} - \partial_{\nu}\omega_{\mu}, \ \ \ \ \ 
	\vec{R}_{\mu\nu} = \partial_{\mu}\vec{\rho}_{\nu} - \partial_{\nu}\vec{\rho}_{\mu},
	\label{eq:covariant-derivative}
\ene
where $\sigma$ and $\omega$ are isoscalar mesons, while ${\vec \pi}$ and ${\vec \rho\,}$ are isovector mesons.   
The meson masses are respectively chosen as 
$m_{\sigma}=550$ MeV, $m_{\omega}=783$ MeV, $m_{\pi}=138$ MeV and $m_{\rho}=770$ MeV.  

The interaction Lagrangian is given by
\bg
	\mathcal{L}_{int} &=& \sum_{B}\bar{\psi}_{B}\left[g_{\sigma B} \sigma
	-g_{\omega B}\gamma_{\mu}\omega^{\mu} 
	+ \frac{f_{\omega B}}{2\mathcal{M}}\sigma_{\mu\nu}\partial^{\nu}\omega^{\mu}\right. \nonumber \\ 
	&-& \left. g_{\rho B}\gamma_{\mu}\vec{\rho}^{\,\mu}\cdot\vec{I}_B
	+ \frac{f_{\rho B}}{2\mathcal{M}}\sigma_{\mu\nu}\partial^{\nu}\vec{\rho}^{\,\mu}\cdot\vec{I}_B
	- \frac{f_{\pi B}}{m}\gamma_{5}\gamma_{\mu}\partial^{\mu}\vec{\pi}\cdot\vec{I}_B \right]\psi_{B},
	\label{eq:Lagrangian-interaction}
\en
where $\vec{I}_B$ is the isospin matrix for baryon $B$ (we set $\vec{I}_B=0$ when $B$ is iso-singlet) 
and the common, scale mass $\mathcal{M} \, (m)$ is taken to be the free nucleon (pion) mass~\citep{esc08}. 
The $\sigma$-, $\omega$-, $\rho$-, $\pi$-$B$ coupling constants are respectively denoted by 
$g_{\sigma B}$, $g_{\omega B}$, $g_{\rho B}$ and $f_{\pi B}$, 
while $f_{\omega B}$ and $f_{\rho B}$ are the isoscalar-tensor and isovector-tensor coupling constants. 
In RHF approximation, the meson field are replaced by the constant mean-field values: $\bar{\sigma}$, $\bar{\omega}$ and $\bar{\rho}$ (the $\rho^0$ field).  
Note that the mean-field value of the pion vanishes.  

In the present calculation, we add non-linear (NL) terms of the $\bar{\sigma}$, 
\bge
	U(\bar{\sigma}) = \frac{g_{2}}{3}\bar{\sigma}^{3} + \frac{g_{3}}{4}\bar{\sigma}^{4}  , 
	\label{eq:scalar-self-interaction}
\ene
to the Lagrangian density, 
because the EoS given by the {\it naive} QHD is too hard, i.e., the incompressibility, $K_v$, for symmetric nuclear matter is calculated to be $\sim 550$ MeV, 
which is apparently too large in comparison with the experimental value, $210-300$ MeV \citep{danielewicz}.  
Here, $g_2$ and $g_3$ are additional coupling constants.  We call this model QHD+NL. 

Although the baryons are assumed to be structureless in QHD, the effect of the size can be included if we introduce a form factor at each interaction vertex. 
In fact, at the Hartree level, it is unnecessary to include it because the momentum transfer in the meson exchange between two baryons vanishes. 
However, because the exchanged momentum can be finite in the Fock term, the effect of form factor may become significant as the density grows up.  
We choose a dipole function as the form factor:
\bge
 F_{i}({p}^{2}) = 
		\frac{1}{\left(1- p^{2}/\Lambda_{i}^{2}\right)^{2}} , \label{formfactor}
\ene
where $p^\mu$ is the (four) momentum transfer, $\Lambda_{i}$ is a cutoff parameter and $i$ specifies the interaction vertex  
which will be discussed below (see also Table~\ref{tab:BSE}). 
Note that the effect of form factor diminishes with increasing $\Lambda_{i}$. 
To study the effect of baryon size, in the interaction Lagrangian density, we replace all coupling constants (except for $g_2$ and $g_3$ in Eq.(\ref{eq:scalar-self-interaction})) with 
those multiplied by the form factor, Eq.~(\ref{formfactor}).  

To sum up all orders of the tadpole (Hartree) and exchange (Fock) diagrams in the baryon Green's function, $G_B$, 
we use the Dyson's equation 
\bge
	G_{B}(k) = G_{B}^{0}(k) + G_{B}^{0}(k)\Sigma_{B}(k)G_{B}(k) , 
	\label{eq:Dyson-equation}
\ene
where $k^\mu$ is the four momentum of baryon, $\Sigma_{B}$ is the baryon self-energy and $G_{B}^{0}$ is the Green's function in free space. 
In a matter, 
the baryon self-energy is generally written as
\bge
	\Sigma_{B}(k) = \Sigma_{B}^{s}(k) - \gamma_{0}\Sigma_{B}^{0}(k) 
	+ (\vec{\gamma}\cdot\hat{k})\Sigma_{B}^{v}(k),
	\label{eq:baryon-self-engy}
\ene
with $\hat{k}$ the unit vector along the (three) momentum $\vec{k}$ and $\Sigma_{B}^{s(0)[v]}$ 
the scalar part (the time component of the vector part) [the space component of the vector part] of the self-energy. 
Furthermore, the effective baryon mass, momentum and energy in a matter are respectively defined by 
\bg
	M_{B}^{\ast}(k) &=& M_{B} + \Sigma_{B}^{s}(k),\\
	k_{B}^{\ast\mu} &=& (k_{B}^{\ast0},\vec{k}_{B}^{\ast}) 
	= (k^{0}+\Sigma_{B}^{0}(k),\vec{k}+\hat{k}\Sigma_{B}^{v}(k)) ,\\
	E_{B}^{\ast}(k) &=& \left[\vec{k}_{B}^{\ast2}+M_{B}^{\ast2}(k)\right]^{1/2}.
	\label{eq:auxiliary-quantity}
\en

The baryon self-energies in Eq.~(\ref{eq:baryon-self-engy}) are then calculated by \citep{bouyssy,miyatsu2012}
\bg
	\Sigma_{B}^{s}(k) &=& -g_{\sigma B} \bar{\sigma}
	+ \sum_{B^{\prime}, i} \frac{(I_{BB^{\prime}}^{i})^2}{(4\pi)^{2}k}\int_{0}^{k_{F_{B^{\prime}}}} dq \, q
	\left[\frac{M_{B^{\prime}}^{\ast}(q)}{E_{B^{\prime}}^{\ast}(q)}B_{i}(k,q)
	+ 
	\frac{q_{B^{\prime}}^{\ast}}{2E_{B^{\prime}}^{\ast}(q)}D_{i}(q,k)\right],
	\label{eq:BSE-scalar} \\
	\Sigma_{B}^{0}(k) &=& -g_{\omega B}\bar{\omega}-g_{\rho B}({\vec I}_{B})_3\bar{\rho}
	-\sum_{B^{\prime}, i} \frac{(I_{BB^{\prime}}^{i})^2}{(4\pi)^{2}k} \int_{0}^{k_{F_{B^{\prime}}}}dq \, q
	A_{i}(k,q),
	\label{eq:BSE-time} \\
	\Sigma_{B}^{v}(k) &=& \sum_{B^{\prime}, i} \frac{(I_{BB^{\prime}}^{i})^2}{(4\pi)^{2}k} 
	\int_{0}^{k_{F_{B^{\prime}}}}dq \, q
	\left[\frac{q_{B^{\prime}}^{\ast}}{E_{B^{\prime}}^{\ast}(q)}C_{i}(k,q)
	+ 
	\frac{M_{B^{\prime}}^{\ast}(q)}{2E_{B^{\prime}}^{\ast}(q)}D_{i}(k,q)\right],
	\label{eq:BSE-vector}
\en
where $k_{F_B}$ is the Fermi momentum for baryon $B$ 
and the factor, $I_{BB^{\prime}}^{i}$, is the isospin weight at the meson-$BB^\prime$ vertex 
in the Fock diagram.  In the present calculation, we ignore the retardation effect 
(i.e., the energy transfer between two interacting baryons vanishes)\footnote{In this approximation, the form factor is given as $F_{i}({p}^{2}) \to  
		\left(1+ {\vec p\,}^{2}/\Lambda_{i}^{2}\right)^{-2}$, where ${\vec p}$ is the (three) momentum transfer.}, 
which gives at most a few per cent contribution to the self-energy \citep{serot1984}. 
The index $i$ in the sum and the functions $A_{i}$, $B_{i}$, $C_{i}$ and $D_{i}$ 
in Eqs.~(\ref{eq:BSE-scalar})-(\ref{eq:BSE-vector}) are explicitly specified in Table~\ref{tab:BSE}, 
in which the following functions are used\footnote{In the loop integral, we have removed the so-called contact interaction term \citep{bouyssy,krein}.}:
\bg
	\Theta_{i}(k,q) &=& \frac{\Lambda_{i}^{8}}{(m_{i}^{2}-\Lambda_{i}^{2})^{4}} 
	 \left(\ln\left[\frac{M_i^{+}(k,q)}{M_i^{-}(k,q)}\frac{L_i^{-}(k,q)}{L_i^{+}(k,q)}\right]
		+ \sum_{n=1}^{3}\left(m_{i}^{2}-\Lambda_{i}^{2}\right)^{n}N_{i}^{n}(k,q)\right) , \\
	\Phi_{i}(k,q) &=& \frac{1}{4kq}
		\left[\left(k^{2}+q^{2}+m_{i}^{2}\right)\Theta_{i}(k,q)
		-\Lambda_{i}^{8}N_{i}^{3}(k,q)\right] ,\\
	\Psi_{i}(k,q) &=& \left(k^{2}+q^{2}-m_{i}^{2}/2\right)\Phi_{i}(k,q)
		-kq\Theta_{i}(k,q) + \Omega_{i}(k,q) , \\
	\Pi_{i}(k,q) &=& \left(k^{2}+q^{2}\right)\Phi_{i}(k,q)
		-kq\Theta_{i}(k,q) + \Omega_{i}(k,q) , \\
	\Gamma_{i}(k,q) &=& \left[k\Theta_{i}(k,q)-2q\Phi_{i}(k,q)\right] ,
\en
where
\bg
	\Omega_{i}(k,q) &=& \frac{\Lambda_{i}^{8}}{4kq}
		\left[N_{i}^{2}(k,q)+\left(k^{2}+q^{2}+\Lambda_{i}^{2}\right)N_{i}^{3}(k,q)\right] ,  \\
		L_{i}^{\pm}(k,q) &=& \Lambda_{i}^{2} + (k \pm q)^{2}, \\
		M_{i}^{\pm}(k,q) &=& m_{i}^{2} + (k \pm q)^{2}, \\
		N_{i}^{n}(k,q) &=& \frac{(-1)^{n}}{n}
		\left(\left[L_{i}^{+}(k,q)\right]^{-n}-\left[L_{i}^{-}(k,q)\right]^{-n}\right) . 
\en
%

The mean-field values of $\bar{\omega}$ and $\bar{\rho}$ are respectively given by the usual forms 
\bge
	\bar{\omega} = \sum_{B}\frac{g_{\omega B}}{m_{\omega}^{2}}\rho_{B}, \ \ \ \ \ 
	\bar{\rho} = \sum_{B}\frac{g_{\rho B}}{m_{\rho}^{2}}({\vec I}_{B})_3\rho_{B},
	\label{eq:rho-field}
\ene
where the density of baryon $B$ is $\rho_{B} = (\frac{2J_{B}+1}{6\pi^{2}}){k^{3}_{F_{B}}}$ with $J_B$ the spin of $B$. 

On the other hand, combining with Eqs.~(\ref{eq:BSE-scalar})-(\ref{eq:BSE-vector}), the $\bar{\sigma}$ value should be self-consistently calculated by \citep{weber}
\bge
	\bar{\sigma} = \sum_{B}\frac{g_{\sigma B}}{m_{\sigma}^{2}}\rho_{B}^s
	- \frac{1}{m_{\sigma}^{2}}(g_{2}\bar{\sigma}^{2}+g_{3}\bar{\sigma}^{3}) , 
	\label{eq:sigma-field}  
\ene
where the scalar density of baryon $B$ reads 
\bge
\rho_{B}^s = \frac{2J_{B}+1}{2\pi^{2}}\int_{0}^{k_{F_{B}}}dk~k^{2}\frac{M_{B}^{\ast}(k)}{E_{B}^{\ast}(k)} . 
	\label{eq:baryon-scalar-density} 
\ene
%

The total energy density and pressure can be divided into the hadron and lepton parts, 
namely $\epsilon=\epsilon_{H}+\epsilon_{\ell}$ and $P=P_{H}+P_{\ell}$, 
where the hadron energy density, $\epsilon_{H}$, and pressure, $P_{H}$, include both the baryon and meson contributions.  Then, they are expressed as 
\bge
	\epsilon_{H} = \sum_{B}\frac{2J_{B}+1}{(2\pi)^{3}}\int_{0}^{k_{F_{B}}}d\vec{k}~\left[
	T_{B}(k)+\frac{1}{2}V_{B}(k)\right]
	-\frac{\bar{\sigma}^{3}}{2}\left(\frac{g_{2}}{3}+\frac{g_{3}}{2}\bar{\sigma}\right),
	\label{eq:baryon-engy-density}
\ene
with
\bge
	T_{B}(k) = \frac{M_{B}M_{B}^{\ast}(k)+kk_{B}^{\ast}}{E_{B}^{\ast}(k)} , \ \ \ \ \ 
	V_{B}(k) = \frac{M_{B}^{\ast}(k)\Sigma_{B}^{s}(k)+k_{B}^{\ast}\Sigma_{B}^{v}(k)}{E_{B}^{\ast}(k)}-\Sigma_{B}^{0}(k)
	\label{eq:baryon-engy-density-kinetic-potential} , 
\ene
and 
\bge
	P_{H} = n_{B}^{2}\frac{\partial}{\partial n_{B}}\left(\frac{\epsilon_{B}}{n_{B}}\right)
	\label{eq:baryon-pressure} , 
\ene
where the total baryon density, $n_{B}$, is given by $n_{B} = \sum_{B}\rho_{B}$.  These energy density and pressure give the EoS for the core matter of neutron star.

\subsection{\label{qmc} QMC and CQMC models}

In QHD, $g_{\sigma B}$ is constant at any density because the baryons are point-like objects.  
In contrast, in the QMC and CQMC models, the $\sigma$-$B$ coupling constant, $g_{\sigma B}$, has the $\sigma$-field dependence, 
which reflects the baryon structure variation due to the interaction in matter. However, the medium effect does {\it not} affect the coupling of vector meson to baryon 
in mean-field approximation \citep{saito2007}.  
Furthermore, in a uniformly distributed matter, it may not be necessary to consider the 
internal structure variation of mesons explicitly, because the meson field is no longer {\em dynamical} but just {\em auxiliary}, namely 
the meson field (or the mean-field) can be replaced by the baryon density or the baryon scalar density \citep{saito2010}. 

In the QMC model, the MIT bag model is simply used to describe the baryons, in which three quarks move freely.  
As an example, we consider the nucleon mass in matter.  When we calculate the mass in RH approximation, it is given as a 
function of the scalar mean-field
\bge
M_N^{\ast}({\bar \sigma}) \equiv M_N - g_{\sigma N}({\bar \sigma}) {\bar \sigma} , 
\label{nuclm0}
\ene
because the scalar field couples to a quark in the nucleon and modifies the quark mass in matter.  Here, $g_{\sigma N}({\bar \sigma})$ is the $\sigma$-$N$ coupling constant, 
which depends on the $\sigma$ field, and will be discussed further below. 

Because the value of ${\bar \sigma}$ is small at low density, the effective nucleon mass in matter, $M_N^{\ast}$, can be expanded in terms of ${\bar \sigma}$ as 
\bge
M_N^{\ast}({\bar \sigma}) = M_N + \left( \frac{\partial M_N^{\ast}}{\partial {\bar \sigma}} 
\right)_{{\bar \sigma}=0} {\bar \sigma} + \frac{1}{2} \left( \frac{\partial^2 M_N^{\ast}}
{\partial {\bar \sigma}^2} \right)_{{\bar \sigma}=0} {\bar \sigma}^2 + \cdots . 
\label{nuclm}
\ene
In the QMC model, the interaction Hamiltonian between the nucleon and the 
$\sigma$ field at the quark level is given by $H_{int} = - 3 g_{\sigma}^q \int_V d{\vec r} \ {\bar \psi}_q \sigma \psi_q$, 
where $g_{\sigma}^q$ is the $\sigma$-quark coupling constant, $\psi_q$ is the quark wave function and $V$ is the nucleon volume, and the derivative of 
$M_N^{\ast}$ with respect to ${\bar \sigma}$ is given by
\bge
\left( \frac{\partial M_N^{\ast}}{\partial {\bar \sigma}} \right) 
= -3g_{\sigma}^q \int_V d{\vec r} \ \ {\bar \psi}_q \psi_q 
\equiv -3g_{\sigma}^q S_N({\bar \sigma}) . \label{deriv}
\ene
Here we have defined the quark-scalar density in the nucleon, $S_N({\bar \sigma})$,
which is itself a function of the scalar field, by 
Eq.(\ref{deriv}). 
Because of a negative value of 
$\left( \frac{\partial M_N^{\ast}}{\partial {\bar \sigma}} \right)$, 
the nucleon mass decreases in matter at low density.  

Moreover, we define the scalar-density ratio (or the scalar polalizability), $S_N({\bar \sigma})/S_N(0)$, 
to be $C_N({\bar \sigma})$ and the $\sigma$-$N$ coupling constant at ${\bar \sigma} = 0$ 
to be $g_{\sigma N}$ (i.e., $g_{\sigma N} \equiv g_{\sigma N}({\bar \sigma}=0)$):
\bge
C_N({\bar \sigma}) = S_N({\bar \sigma})/S_N(0) \ \ \mbox{and} \ \ 
g_{\sigma N} = 3g_{\sigma}^q S_N(0) . \label{cn}
\ene
Comparing with Eq.(\ref{nuclm0}), we find that 
\bge
\left( \frac{\partial M_N^{\ast}}{\partial {\bar \sigma}} \right) 
= -g_{\sigma N} C_N({\bar \sigma}) = - \frac{\partial}{\partial {\bar \sigma}}
\left[ g_{\sigma N}({\bar \sigma}) {\bar \sigma} \right],
\label{deriv2}
\ene
and that the nucleon mass is 
\bge
M_N^{\ast} = M_N - g_{\sigma N} {\bar \sigma} - \frac{1}{2} g_{\sigma N} 
C_N^\prime(0) {\bar \sigma}^2 + \cdots . 
\label{nuclm2}
\ene
In general, $C_N$ is a decreasing function because the quark in matter becomes 
more relativistic than in free space.  Thus, $C_N^\prime(0)$ takes a 
negative value. If the nucleon were structureless, $C_N$ would not depend on 
the scalar field, that is, $C_N$ would be constant ($C_N=1$).  Therefore, 
only the first two terms in the right hand side of Eq.(\ref{nuclm2}) remain, 
which is exactly the same as the equation for the effective nucleon 
mass in QHD.  This decrease in $C_N$
constitutes a new saturation mechanism for nuclear matter \citep{guichon1988}, and very much reduces the nuclear incompressibility $K_v$, 
compared with the value of QHD.  Therefore, in the QMC model, it is not necessary to include the non-linear terms of ${\bar \sigma}$.  

In actual numerical calculations, we found that 
the scalar-density ratio, $C_N(\sigma)$, decreases linearly 
(to a very good approximation) with $g_{\sigma N} {\bar \sigma}$ \citep{guichon1996,saito1997}.  Then, it is very useful to have a simple 
parameterization for $C_N$: 
\bge
C_N({\bar \sigma}) = 1 - a_N \times (g_{\sigma N} {\bar \sigma}) . 
\label{paramCN}
\ene
As a practical matter, it is easy to solve Eq.(\ref{deriv2}) for 
$g_{\sigma N}({\bar \sigma})$ in the case where $C({\bar \sigma})$ is linear in $g_{\sigma N} {\bar \sigma}$, as in 
Eq.(\ref{paramCN}).  Then one finds 
\bge
M^{\ast}_N =  M_N - g_{\sigma N}({\bar \sigma}) {\bar \sigma} = M_N - g_{\sigma N} \left[ 1 - \frac{a_N}{2} (g_{\sigma N}  
{\bar \sigma}) \right] {\bar \sigma} , 
\label{mstaR}
\ene
so that the effective $\sigma$-N coupling constant, $g_{\sigma N}({\bar \sigma})$, 
decreases at half the rate of $C_N({\bar \sigma})$.  

For the octet baryons, this linear approximation is also applicable to the scalar-density ratio and the coupling constant in a nuclear medium: 
\bg
C_B({\bar \sigma}) &=& 1 - a_B \times (g_{\sigma N} {\bar \sigma}) , \label{scal-den} \\
g_{\sigma B}(\bar{\sigma}) &=& g_{\sigma B} \left[1-\frac{a_{B}}{2}(g_{\sigma N}\bar{\sigma})\right]  , 
\label{paramCNB}
\en
with $a_B$ a parameter and $g_{\sigma B}$ the $\sigma$-$B$ coupling constant at $n_B =0$.  
This idea leads to a new, simple scaling relationship among the in-medium baryon masses \citep{saito1997}. 

Two quarks in a baryon can interact through the one-gluon exchange (OGE) \citep{guichon2008}.  
In addition to it, the pion-cloud effect, which is governed by chiral symmetry and its small breaking, 
may also affect the properties of baryons in a nuclear medium.  The CQMC model involves both the effects \citep{nagai,miyatsu2009}.  
The hyperfine interaction due to the OGE and pion cloud is very significant in the in-medium baryon spectra.  In particular, at the quark mean-field level, it splits the 
masses of $\Lambda$ and $\Sigma$ in matter, and, as it should be, the $\Lambda$ feels more attractive force than the $\Sigma$ or $\Xi$ does.  

\begin{figure}
\includegraphics[angle=-90,scale=0.8]{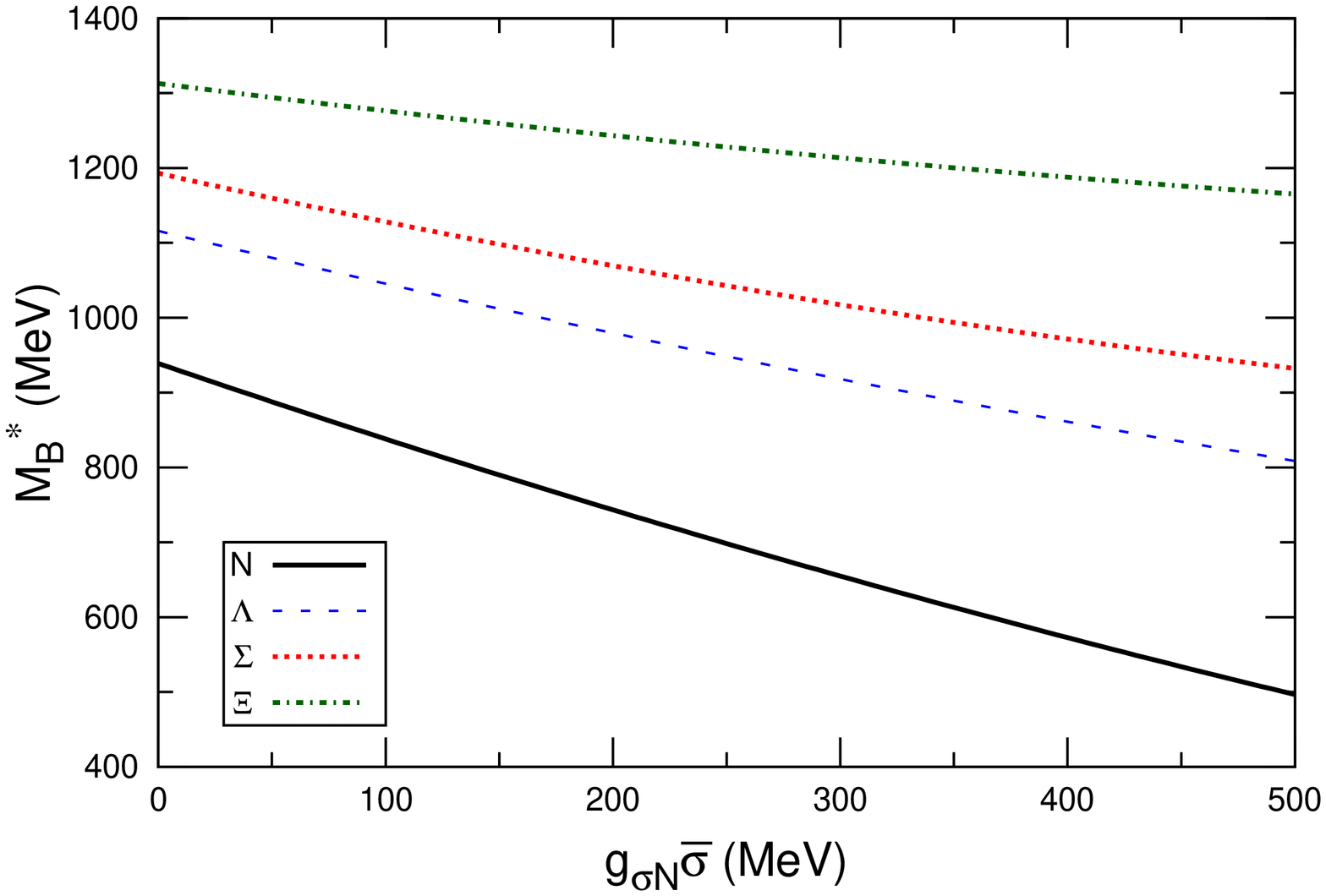}
\caption{The CQMC result for octet baryon masses (as functions of the scalar field $g_{\sigma N}{\bar \sigma}$) in symmetric nuclear matter. 
We note that the difference between the results by the exact calculation and the parameterization, Eq.(\ref{sigma-coupling-const}), is indistinguishable.  
Here, we use the coupling constants, $g_{\sigma B}$, given by SU(6) symmetry, namely 
$g_{\sigma N}=\frac{3}{2}g_{\sigma\Lambda}=\frac{3}{2}g_{\sigma\Sigma}=3g_{\sigma\Xi}$ (see section~\ref{su6}). 
\label{fig:Emass}}
\end{figure}

In the case of the CQMC model, we need two parameters, $a_B$ and $b_B$, to describe the $\sigma$-$B$ coupling constant: 
\bge
	g_{\sigma B}(\bar{\sigma}) = g_{\sigma B}b_{B}\left[1-\frac{a_{B}}{2}(g_{\sigma N}\bar{\sigma})\right],
	\label{sigma-coupling-const}
\ene
while the scalar-density ratio can again be given by Eq.~(\ref{scal-den}).  
This linear approximation is quite accurate up to $\sim 10n_0$. 
The values of $a_B$ and $b_B$ are listed in Table~\ref{tab:parametrizationQMC}.  
Using Eq.(\ref{sigma-coupling-const}), the octet baryon mass in RH approximation is then given by
$M^{\ast}_B =  M_B - g_{\sigma B}({\bar \sigma}) {\bar \sigma}$, which is shown in Fig.~\ref{fig:Emass}. 
If we set $a_{B}=0$ and $b_{B}=1$, $g_{\sigma B}(\bar{\sigma})$ is identical to the $\sigma$-$B$ coupling constant in QHD. 
We should note that the $\sigma$-field dependence in the coupling constant in Eq.~(\ref{sigma-coupling-const}) can also be understood as 
the effect of many-body correlations in nuclear matter \citep{guichon2004,saito2010}

For the RHF calculation in the QMC or CQMC model, we use the formulation developed in section~\ref{rhf}, in which, instead of the constant $g_{\sigma B}$, 
the field-dependent coupling, $g_{\sigma B}(\bar{\sigma})$, in Eq.~(\ref{sigma-coupling-const}) is used at each $\sigma$-$B$ vertex.

\section{Numerical results}

\subsection{\label{su6} SU(6) symmetry}

We focus on how the Fock term including the tensor coupling and the form factor at interaction vertex contribute to the 
EoS for a neutron star.  
There is large ambiguity in determining the meson-hyperon coupling constants.  Then, the SU(6) (quark model) symmetry is often assumed to fix 
the coupling constants 
\citep{reuber,huber}:  
\begin{eqnarray}
g_{\sigma N}&=&\frac{3}{2}g_{\sigma\Lambda}=\frac{3}{2}g_{\sigma\Sigma}=3g_{\sigma\Xi}, \quad \quad 
g_{\omega N}=\frac{3}{2}g_{\omega\Lambda}=\frac{3}{2}g_{\omega\Sigma}=3g_{\omega\Xi},\label{sigma-omega-c.c.}\\
g_{\rho N}&=&\frac{1}{3}g_{\omega N}=\frac{1}{2}g_{\rho \Sigma}=g_{\rho\Xi}, \quad \quad g_{\rho\Lambda}=0,\label{rho-c.c.}\\
f_{\pi N}&=&\frac{5}{4}f_{\pi\Sigma}=-5f_{\pi\Xi}, \quad \quad  f_{\pi\Lambda}=0 ,  \label{pi-c.c.} \\ 
\kappa_{\omega N} &=& \frac{f_{\omega N}}{g_{\omega N}}=-0.12, \quad \quad \kappa_{\omega \Lambda} = \frac{f_{\omega \Lambda}}{g_{\omega \Lambda}}=-1 , \label{f-ov-g-1} \\
\kappa_{\omega \Sigma} &=& \frac{f_{\omega\Sigma}}{g_{\omega\Sigma}}=1+2\kappa_{\omega N} = 0.76, \\
\kappa_{\omega \Xi} &=& \frac{f_{\omega\Xi}}{g_{\omega\Xi}}=-2-\kappa_{\omega N} = -1.88 , \label{f-ov-g-2} \\
\kappa_{\rho N} &=& \frac{f_{\rho N}}{g_{\rho N}}=3.7, \quad \quad \kappa_{\rho \Lambda} = \frac{f_{\rho \Lambda}}{g_{\rho \Lambda}}= 0 , \label{f-ov-g-3} \\
\kappa_{\rho \Sigma} &=& 	\frac{f_{\rho \Sigma}}{g_{\rho \Sigma}}=-\frac{3}{5}+\frac{2}{5}\kappa_{\rho N} = 0.88 , \\
\kappa_{\rho \Xi} &=& \frac{f_{\rho \Xi}}{g_{\rho \Xi}}=-\frac{6}{5}-\frac{1}{5}\kappa_{\rho N} = -1.94 ,  \label{f-ov-g-4} 
\end{eqnarray}
where we take $f_{\pi N}^2 /4\pi = 0.08$, and $\kappa$ is defined by the ratio of the tensor ($f$) to vector ($g$) coupling constants. 

To see the effects of the Fock term and form factor,  we consider the following cases: \\
(1) RH calculation in QHD-NL, \\
(2) RHF calculation (no tensor coupling, no form factor) in QHD-NL, \\
(3) RHF calculation with tensor coupling (no form factor) in QHD-NL, \\
(4) RHF calculation with tensor coupling and form factor (full calculation) in QHD-NL,  \\
(5) full RHF calculation in QMC, \\
(6) full RHF calculation in CQMC, \\
(7) full RHF calculation with hyperon-potential fit in QHD-NL. \\
For simplicity, in cases of (4) - (7), all cutoff parameters in the form factors are assumed to be the common value, $\Lambda_B^2 = 0.71$ GeV$^2$ \citep{henley}. 
Once $g_{\sigma N}$ is fixed, the $\sigma$-$Y$ couplings, $g_{\sigma Y}$, are determined through the relation Eq.(\ref{sigma-omega-c.c.}). 
In cases of (1) - (6), we use such values for $g_{\sigma Y}$. 
On the other hand, the recent analyses of hypernuclei and hyperon production reactions infer that 
the $\Lambda$,  $\Sigma$ or $\Xi$, respectively, feels the potential, $U_{\Lambda, \ \Sigma, \ \Xi} \simeq -30, +30, -15$ MeV, in a nuclear medium \citep{schaffner,ishizuka}.  
Therefore, in case (7), instead of Eq.(\ref{sigma-omega-c.c.}), we determine $g_{\sigma \Lambda}$, $g_{\sigma \Sigma}$ and $g_{\sigma \Xi}$ so as to fit 
the suggested potential depths at $n_0$.  Apparently, this leads to the breaking of SU(6) symmetry in the coupling constants. 

In Table~\ref{tab:c.c.s}, 
we list the coupling constants and several properties of symmetric nuclear matter at $n_0$. 
We have four, free parameters in QHD-NL: $g_{\sigma N}$, $g_{\omega N}$, $g_2$ and $g_3$. 
The first two parameters, $g_{\sigma N}$ and $g_{\omega N}$, are adjusted so as to reproduce the nuclear saturation energy ($-15.7$ MeV) at $n_0$ in symmetric nuclear matter.  
The last two constants, $g_2$ and $g_3$, are determined so as to generate the incompressibility $K_v$ (=253 MeV) and the effective nucleon mass ($M_N^{\ast}/M_N=0.79$) 
calculated by the QMC model (case (5)).   
For case (7), the coupling constants except for $g_{\sigma Y}$ are identical to those in case (4), and we use 
$g_{\sigma \Lambda}^2/4\pi = 1.97$, $g_{\sigma \Sigma}^2/4\pi = 0.38$ and $g_{\sigma \Xi}^2/4\pi = 0.52$, which produce the required hyperon-potential depths. 
In the QMC and CQMC models (cases (5) and (6)), we have only two parameters, 
$g_{\sigma N}$ and $g_{\omega N}$, which are again determined so as to satisfy the nuclear saturation condition.  We note that the values of $K_v$, $a_4$ and $M_B^{\ast}$ 
calculated by the QMC are 
different from those by the CQMC, which is caused by the hyperfine interaction due to the gluon and pion exchanges between quarks, as discussed in section \ref{qmc}. 

Comparing cases (1) - (3), we can see that, because the Fock terms, especially the terms involving the tensor couplings, contribute as an attractive force, 
the coupling constants, $g_{\sigma N}$ and $g_{\omega N}$, are reduced in cases (2) and (3) to satisfy the saturation condition for nuclear matter. 
In case (3), the Fock contribution at $n_0$ reaches about 35 (25)\% of the scalar (vector) self-energy.  In contrast, the inclusion of form factors  reduces  
the Fock contribution, which is shown in the contents of the self-energies in case (4), and $g_{\sigma N}$ and $g_{\omega N}$ are accordingly enhanced. 
Because of the quark and gluon degrees of freedom, the scalar and vector self-energies in the CQMC model (case (6)) are more enhanced (deeper) than in QHD-NL, 
which generates the lower effective nucleon mass in matter and the larger nuclear incompressibility.  

\begin{figure}
\includegraphics[angle=-90,scale=0.6]{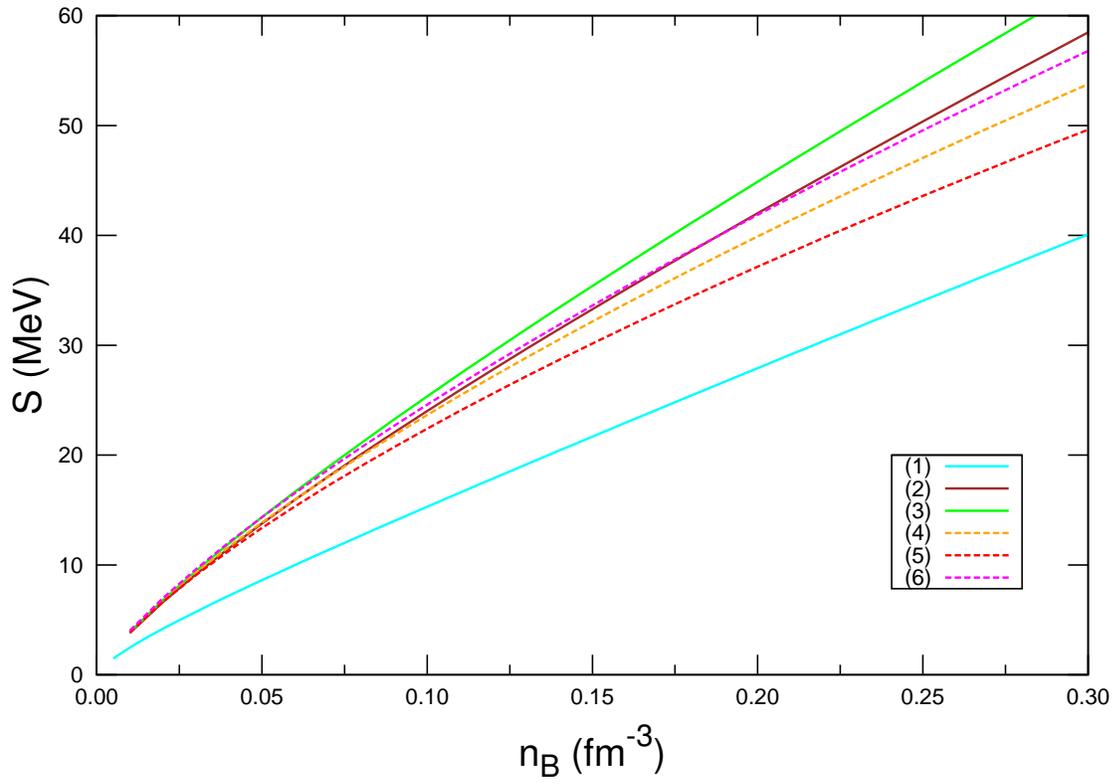}
\caption{Symmetry energy $a_4$ in SU(6) symmetry.  \label{fig:a4-su6}}
\end{figure}

We should notice that the Fock term plays a very important role in the symmetry energy, $a_4$.  
The value of $a_4$ in case (1) is very small, while, the Fock term enhances $a_4$ very much (see cases (2) - (6)). 
We present the density dependence of $a_4$ in Fig.~\ref{fig:a4-su6}.  
In the RH calculation (case (1)), it is necessary to enhance the $\rho$-$N$ vector coupling by about $45\%$ {\em by hand} to obtain the observed symmetry energy, 
$a_4 \simeq 30$ MeV. 

\begin{figure}
\includegraphics[angle=-90,scale=0.6]{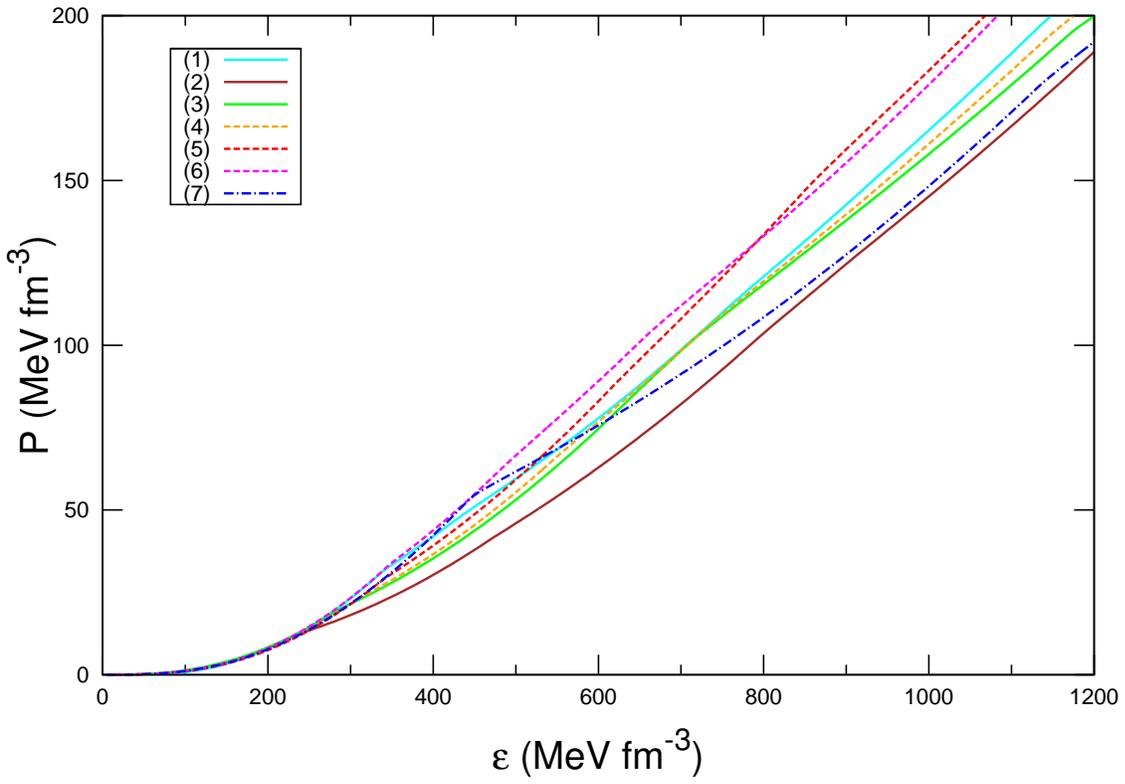}
\caption{Equation of state (pressure versus energy density) in SU(6) symmetry.  \label{fig:EoS-su6}}
\end{figure}

\begin{figure}
\epsscale{0.80}
\plotone{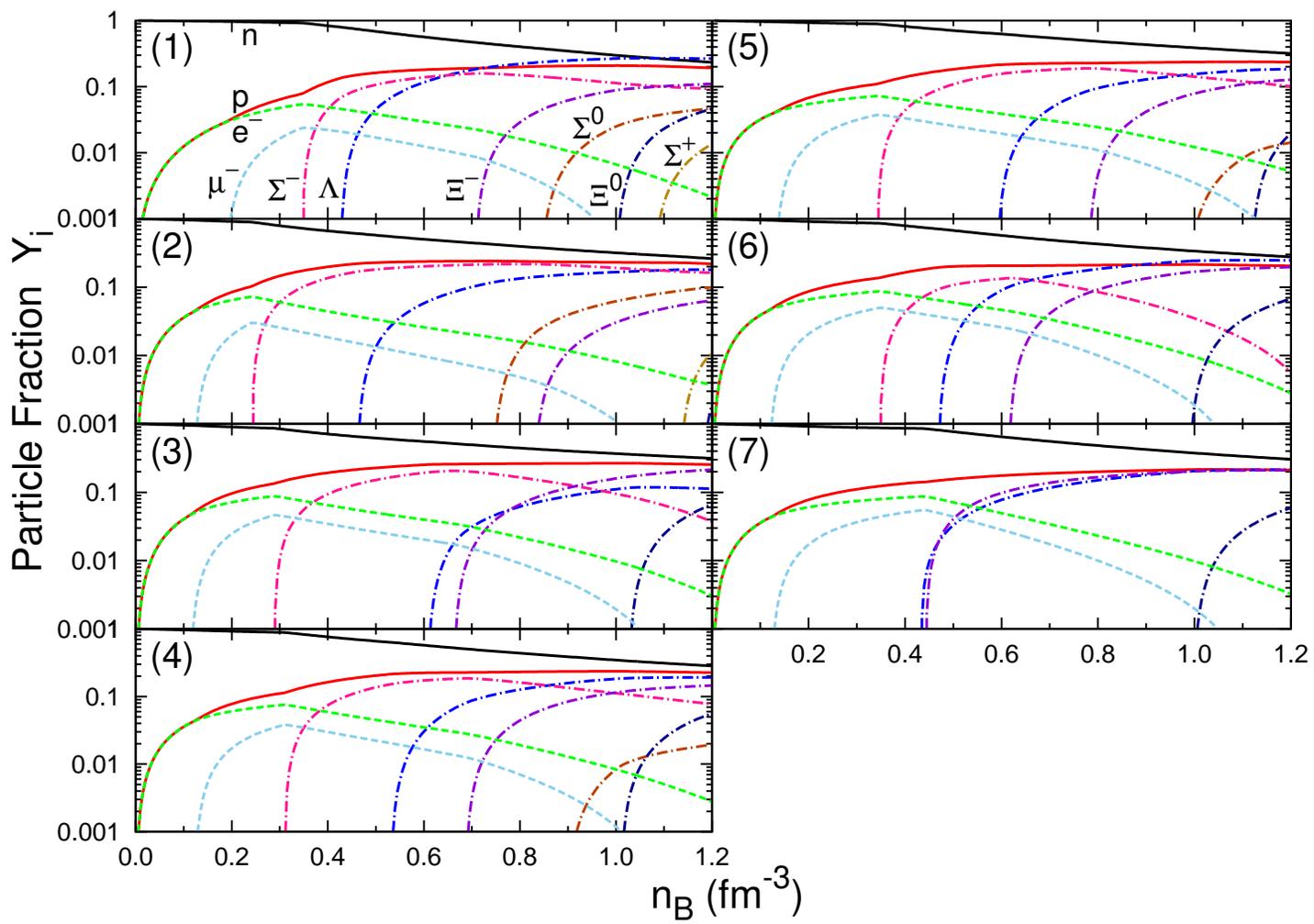}
\caption{Particle fractions, $Y_i$, for cases (1) - (7). \label{fig:pf-su6}}
\end{figure}

In Figs.~\ref{fig:EoS-su6} and \ref{fig:pf-su6}, we respectively show the equation of state and the particle fractions, $Y_i$ ($i = n, p, \cdots$), in the core matter of neutron star, 
which are calculated under the conditions of the charge neutrality and the $\beta$-equilibrium in weak interaction. 
As expected, the hard EoS can be given by the QMC or CQMC model, while the softest one is generated by case (2).  In Fig.~\ref{fig:pf-su6}, we can see 
all octet baryons in the RH result (case (1)).  The Fock terms (without the tensor couplings) push upward the threshold densities for hyperons 
(except for $\Sigma^-$ and $\Sigma^0$), see case (2).  Furthermore, the inclusion of the tensor coupling (case (3)) makes the threshold densities 
of $\Lambda$ and $\Sigma^0$ very high (actually, $\Sigma^0$ disappear in the figure), while that of the $\Xi^-$ is somewhat pulled downward.  
In case (4), we can see that such tensor effect is weakened by the inclusion of form factors.  In case (5), comparing with case (4), the thresholds of hyperons 
are again pushed upward, which is caused by the effect of baryon internal structure variation in matter.  The difference between cases (5) and (6) is generated by  
the hyperfine interaction between quarks in octet baryons. In case (7), as expected, the $\Sigma$ hyperons do not appear at all below $n_B = 1.2$ fm$^{-3}$, which is 
caused by the adjustment of the coupling constant $g_{\sigma \Sigma}$ to fit the repulsive potential depth, $U_\Sigma \simeq + 30$ MeV. 

\begin{figure}
\includegraphics[angle=-90,scale=1.0]{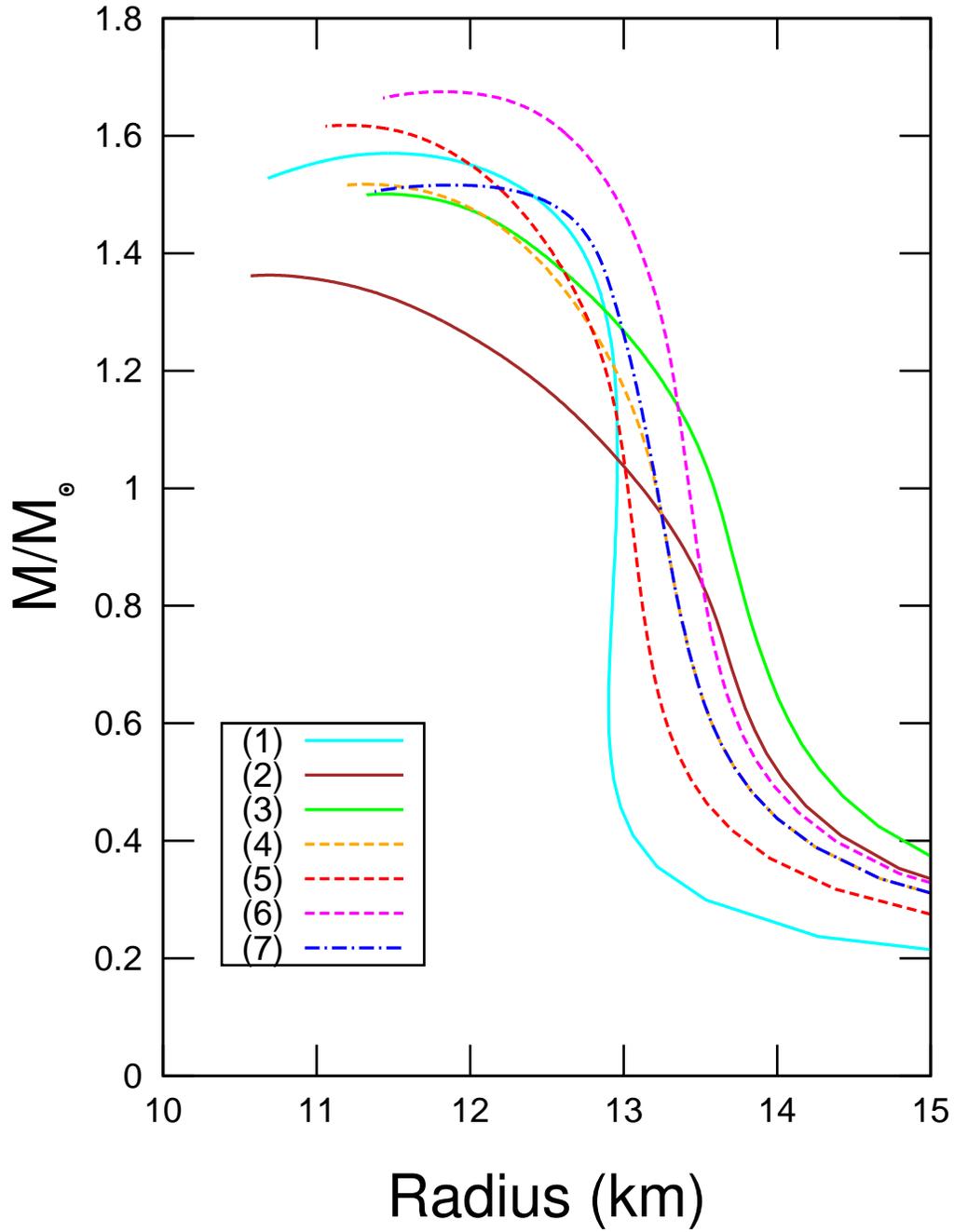}
\caption{Neutron star mass versus radius in SU(6) symmetry.  \label{fig:stellar-su6}}
\end{figure}

We respectively present the neutron-star mass as a function of the radius and the properties of neutron star in Fig.~\ref{fig:stellar-su6} and Table~\ref{tab:n.s.mass-su6}.  
To calculate the neutron-star mass, we solve the Tolman-Oppenheimer-Volkoff (TOV) equation \citep{tolman,oppen}.  
As expected from Fig.~\ref{fig:EoS-su6}, the CQMC model gives the 
heaviest mass.  In contrast, although the mass is low, the case (2) gives the neutron star whose central density is very high, namely it is very compact.  In SU(6) symmetry, 
the maximum neutron-star mass gradually increases from case (2) to (4).  

\subsection{\label{su3} SU(3) symmetry}

To study the role of hyperons on the properties of neutron star, it is very important to extend SU(6) symmetry to the more general SU(3) flavor symmetry \citep{weissenborn2012b}. 
Restricting our interest to three quark flavors, up, down and strange, SU(3) symmetry in flavor space can be regarded as a symmetry group of strong interaction. 
Because we consider only the octet representation for the baryons ($J^P = \frac{1}{2}^+$ octet), the SU(3) invariant Lagrangian can be constructed 
using the matrix representations for the baryons, $B$, and meson nonet (singlet state, $M_1$, and octet state, $M_8$).  The interaction Lagrangian is given as a sum of three 
terms, namely one stemming from the coupling of the meson singlet to the baryon octet ($S$ term) and 
the other two terms from the interaction of the meson octet and 
the baryons -- one being the antisymmetric ($F$) term and the other being the symmetric ($D$) term: 
\bge
\mathcal{L}_{int} = -g_8 \sqrt{2} \left[ \alpha {\rm Tr}([{\bar B}, M_8] B) + (1-\alpha ) {\rm Tr}(\{{\bar B}, M_8\} B) \right] - g_1 \frac{1}{\sqrt{3}} {\rm Tr}({\bar B}B){\rm Tr}(M_1) , 
\label{su3-lag}
\ene
where $g_1$ and $g_8$ are respectively the coupling constants for the meson singlet and octet states, and $\alpha$ ($0 \leq \alpha \leq 1$) denotes the $F/(F+D)$ ratio, namely 
the ratio of the antisymmetric contribution, $F$, to the sum of antisymmetric and symmetric, $D$, ones.  
In SU(3) symmetry, all possible combinations of couplings are then described in terms of four parameters; the singlet and octet coupling constants, $g_1$ and $g_8$, 
$\alpha$ and the mixing angle, $\theta$, relating the physical isoscalar mesons to the pure singlet and octet counterparts.  

The extended-soft-core (ESC08) model by the Nijmegen group may at present be the most complete meson-exchange model for the baryon-baryon interaction \citep{esc08}.  
It describes not only the $N$-$N$ but also the $Y$-$N$ and $Y$-$Y$ interactions in terms of the meson exchange based on SU(3) flavor symmetry.  In this subsection, 
we calculate the EoS using the meson-baryon coupling constants provided by the ESC08 model.  

In this model, for example, the couplings of $\omega$-meson to baryons are related through  
\bge
\frac{g_{\omega \Lambda}}{g_{\omega N}} = \frac{1}{1 + \sqrt{3} z \tan \theta_v} , \quad \quad 
\frac{g_{\omega \Xi}}{g_{\omega N}} = \frac{1-\sqrt{3} z \tan \theta_v}{1 + \sqrt{3} z \tan \theta_v} , 
\label{su3-vec}
\ene
where $z=g_8/g_1$.  The isoscalar $\omega$ and $\phi$ meson are described in terms of the pure singlet, $|1 \rangle$, and 
octet, $|8 \rangle$, states as 
\begin{eqnarray}
|\omega \rangle &=& \cos \theta_v |1 \rangle +  \sin \theta_v |8 \rangle ,   \label{mix-1} \\ 
|\phi \rangle &=& - \sin \theta_v |1 \rangle +  \cos \theta_v |8 \rangle ,  \label{mix-2}  
\end{eqnarray}
where the mixing angle is chosen to be $\theta_v = 37.50^\circ$, which is very close to the ideal mixing, $\tan \theta_v^{ideal} = 1/\sqrt{2}$, 
namely $\theta_v^{ideal} = 35.26^\circ$. 
The ratio $z$ is fixed to be $0.1949$ in the ESC08 model, while $z=1/\sqrt{6} = 0.4082$ in SU(6) symmetry.  Thus, the ratio $z$ is much smaller than the 
SU(6) value, which helps to enhance the maximum mass of neutron star \citep{weissenborn2012b}.  
We show the coupling constants (except for $g_{\sigma B}$ and $g_{\omega N}$) and 
the cutoff parameters in the form factors in Table~\ref{tab:c.c.esc08}.  
It should be noted that, in the present calculation, we use 
the dipole function for the form factor, Eq.(\ref{formfactor}), while, in the ESC08 model, it is expressed by a gaussian function (for details, see the footnotes of Table~\ref{tab:c.c.esc08}). 

Using these coupling constants and cutoff parameters, we here examine the following four cases: \\
(a) full RH calculation in QHD-NL, \\
(b) full RHF calculation in QHD-NL, \\
(c) full RHF calculation in QMC, \\
(d) full RHF calculation in CQMC.   \\
The properties of symmetric nuclear matter at $n_0$ and 
the rest of coupling constants are shown in Tables~\ref{tab:prop.esc08} and \ref{tab:c.c.sY.esc08}.  
As in the case of SU(6) symmetry, the coupling constants, $g_{\sigma N}$ and $g_{\omega N}$, are determined by the nuclear saturation condition.  
In QHD-NL (cases (a) and (b)), the coupling constants for the non-linear terms, $g_2$ and $g_3$, are again determined 
so as to fit the incompressibility and the effective nucleon mass to the values ($K_v=249$ MeV and $M_N^{\ast}/M_N=0.78$) given by the QMC model (case (c)). 
As discussed in subsection~\ref{su6},
the $\sigma$-$Y$ coupling constant (see column (A) in Table~\ref{tab:c.c.sY.esc08}) is fixed so as to reproduce the observed hyperon-potential depth, $U_Y$.  

\begin{figure}
\includegraphics[angle=-90,scale=0.6]{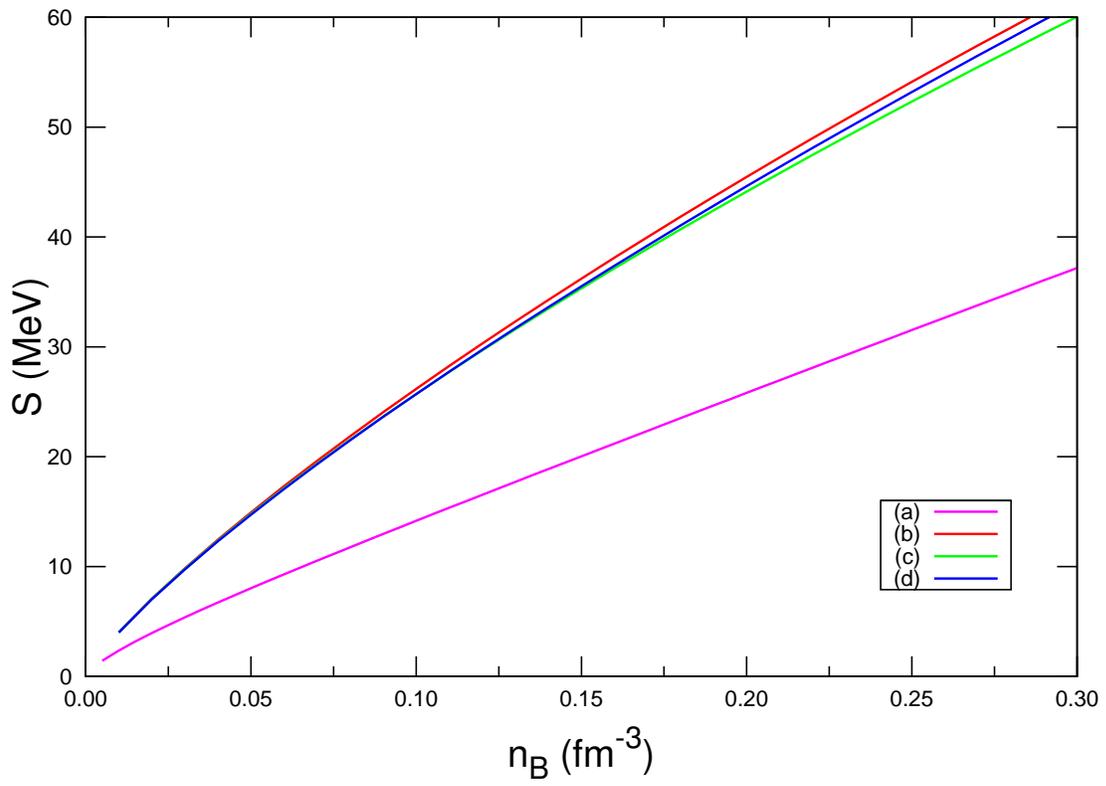}
\caption{Symmetry energy, $a_4$, in the ESC08 model.   \label{fig:a4-su3}}
\end{figure}

\begin{figure}
\includegraphics[angle=-90,scale=0.6]{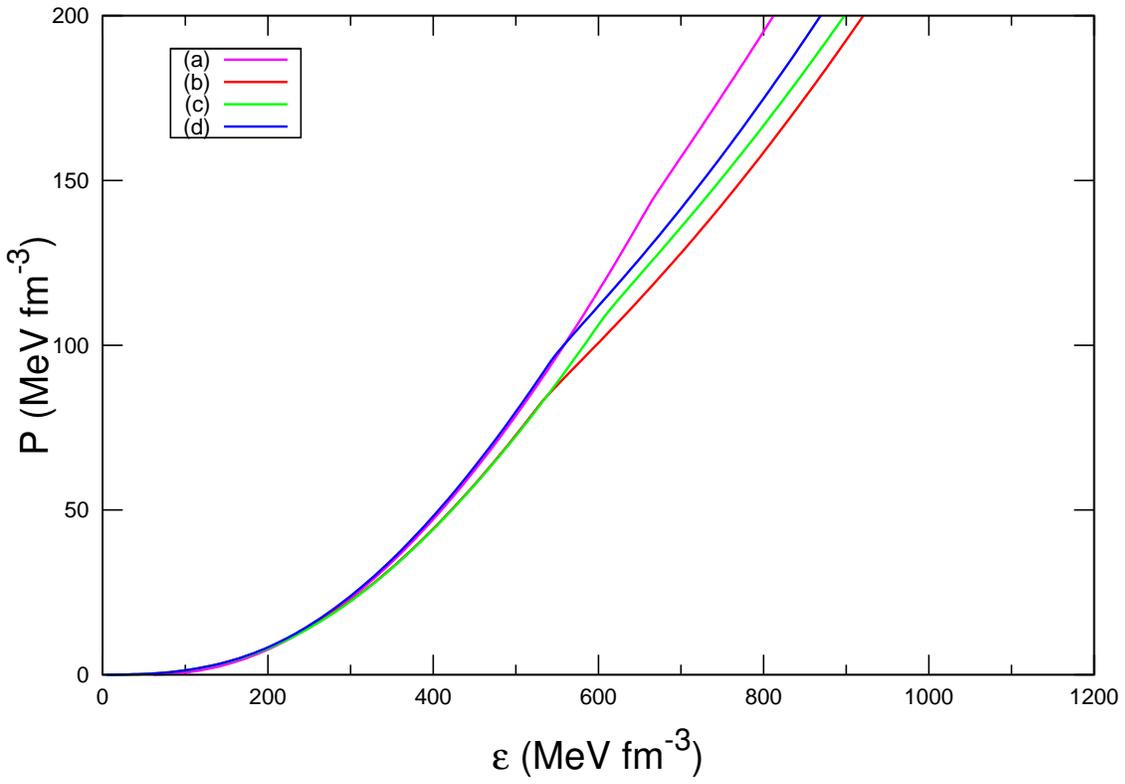}
\caption{Equation of state (pressure versus energy density) in the ESC08 model.  \label{fig:EoS-su3}}
\end{figure}

\begin{figure}
\includegraphics[angle=-90,scale=0.6]{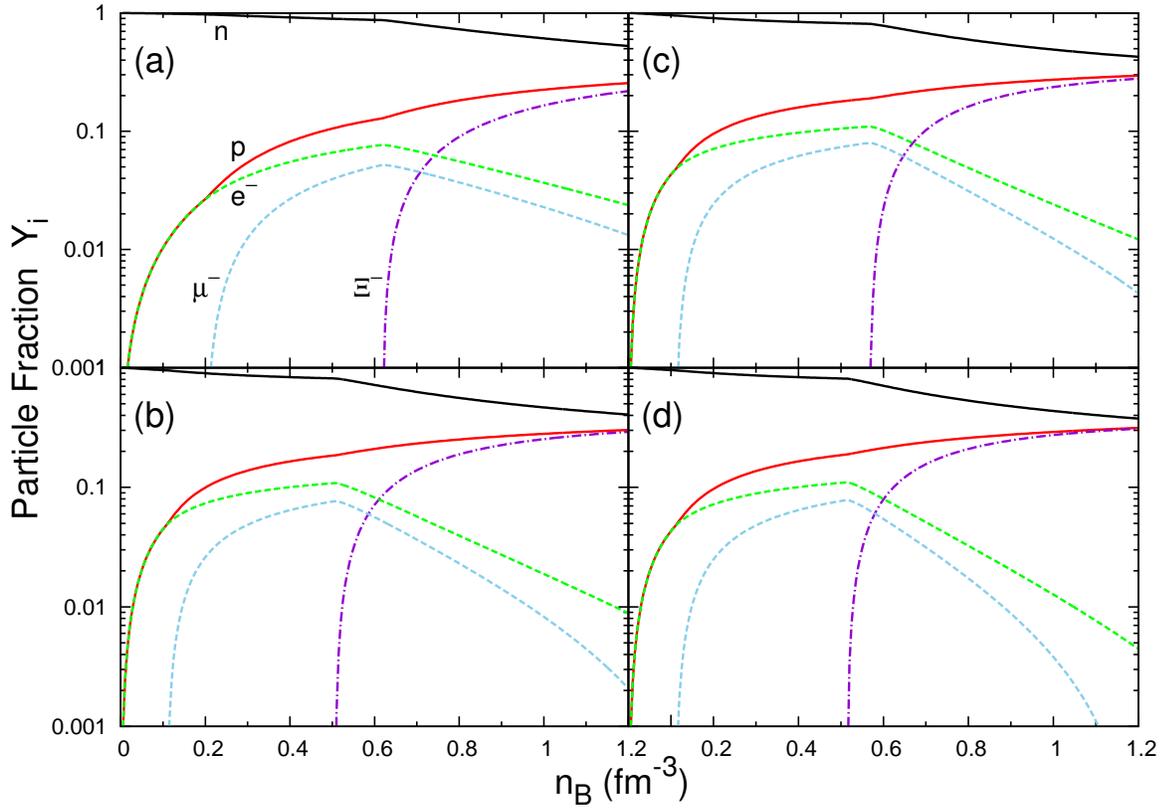}
\caption{Particle fractions, $Y_i$, for cases (a) - (d). \label{fig:pf-su3}}
\end{figure}

We show the symmetry energy, $a_4$, as a function of $n_B$ in Fig.~\ref{fig:a4-su3}. 
It should be again emphasized that in case (a) the symmetry energy is clearly insufficient because of the absence of the Fock term. 
In contrast, $a_4$ is very enhanced in 
cases (b) - (d).  In Figs.~\ref{fig:EoS-su3} and \ref{fig:pf-su3}, we respectively present the equation of state and the particle fractions in the core matter. 
The structure of particle fractions is very simple and impressive in all cases, namely {\it only} the $\Xi^-$ appears below $n_B = 1.2$ fm$^{-3}$. 
In the EoS,  we can see the bend of pressure line around $\varepsilon \simeq 550$ MeV/fm$^3$, which corresponds to the threshold point of the $\Xi^-$ production.  
Beyond the bend, the pressure in the RH calculation (case (a)) is enhanced compared with the 
other cases (b) - (d).  It may be caused by the fact that, in comparison with cases (b) - (d), the production of $\Xi^-$ in case (a) is relatively suppressed 
for high densities, and that the fraction of neutron is thus not reduced much.  
This is also consistent with the fact that the threshold density for the $\Xi^-$ production in case (a) 
is relatively high ($> 0.6$ fm$^{-3}$), comparing with the other cases.  As seen in subsection~\ref{su6}, because the tensor coupling of vector meson pulls downward the 
threshold density of the $\Xi^-$, the reduction of the threshold density in cases (b) - (d) may be originated by the effect of tensor coupling.  

\begin{figure}
\includegraphics[angle=-90,scale=1.0]{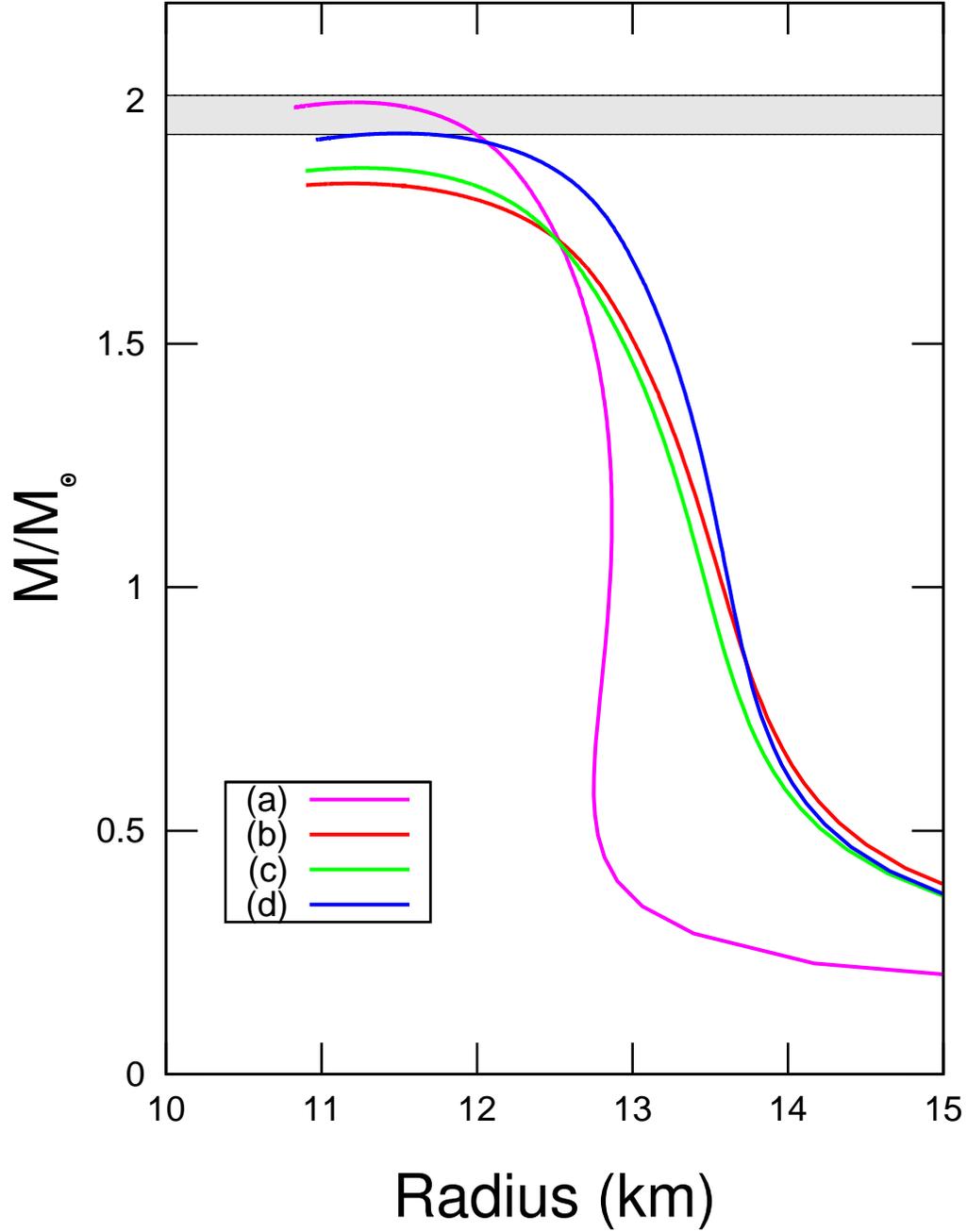}
\caption{Neutron star mass versus radius for cases (a) - (d). 
The shaded area shows the mass of pulsar (PSR J1614-2230), $1.97 \pm 0.04 M_\sun$. \label{fig:stellar-su3}}
\end{figure}

In Fig.~\ref{fig:stellar-su3}, we present the neutron star mass as a function of the star radius.  Because of the hard EoS, the RH calculation gives the heaviest maximum 
mass around $2.0 M_\sun$.  The precise results for the neutron-star properties are presented in column (A) of Table~\ref{tab:n.s.mass-su3}.  The CQMC model gives the 
mass of $1.93 M_\sun$,\footnote{In the previous calculation by \citet{miyatsu2012}, we have reported $M_{max} \simeq 2.02 M_\sun$ in the CQMC model, in which 
the CQMC predicts $K_v = 300$ MeV because the form factor is {\em not} included. 
The difference between the previous and present results is mainly caused by the effect of form factors at interaction vertexes. } 
which is also consistent with the recently observed mass, $1.97 \pm 0.04 M_\sun$ (PSR J1614-2230).  
Comparing the results of cases (b) and (d), we conclude that the change of internal structure of baryons in matter can generate the mass fraction of $\sim 0.1 M_\sun$.  Furthermore, 
because the difference between the QMC and CQMC models is caused by the hyperfine interaction between quarks in a baryon, such interaction is also important and produces 
the mass fraction of $\sim 0.07 M_\sun$. 

We here notice that, because the coupling constants, $g_{\sigma B}$ and $g_{\omega N}$, are determined so as to satisfy the desired conditions, namely the 
saturation condition for symmetric nuclear matter and the observed hyperon-potential depths, SU(3) flavor symmetry in the coupling constants is now partly broken.  
However, the violation of SU(3) symmetry may naturally be expected, because the core matter of neutron star consists of octet baryons whose 
particle fractions are not constant but vary depending on the density.  Furthermore, 
the {\em effective} coupling constants in matter may, in general, vary depending on the surrounding matter, if we treat the matter 
using more sophisticated models such as the (relativistic) Dirac-Brueckner-Hartree-Fock calculations \citep{krastev,dalen}, etc. 

\begin{figure}
\includegraphics[angle=-90,scale=0.6]{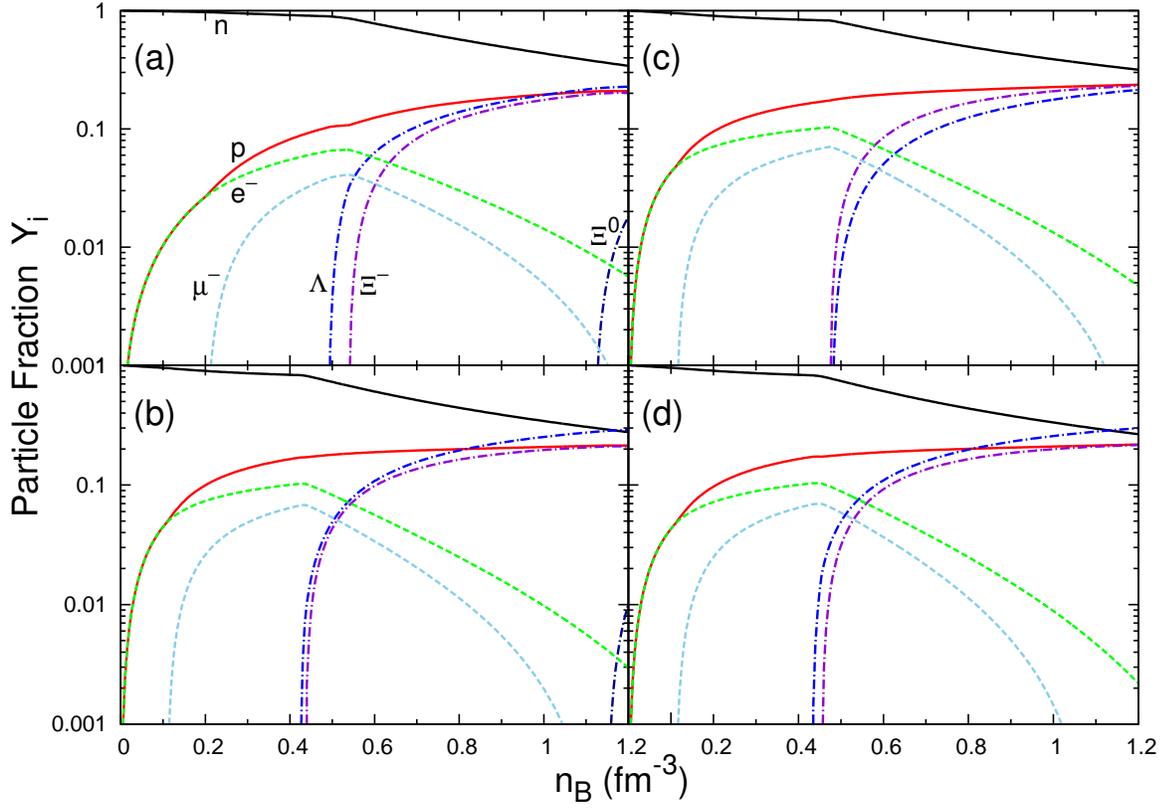}
\caption{Particle fractions, $Y_i$, for cases (a) - (d). The $\omega$-$Y$ coupling constants are reduced to restore the SU(3) relation (for details, see the text). \label{fig:pf-su3-red}}
\end{figure}

However, it may be interesting to examine the neutron-star mass again, restoring SU(3) symmetry.  For example, we try to recalculate the neutron-star mass by reducing the 
$\omega$-$Y$ vector coupling constants by the factor $x(i) = g_{\omega N}^i/g_{\omega N}^0$, where 
$g_{\omega N}^i$ and $g_{\omega N}^0$ are, respectively, the $\omega$-$N$ coupling constants for case ($i $ = a, b, c or d) (see Table~\ref{tab:prop.esc08}) and 
that in free space (i.e., the value given in the ESC08 model).  
We choose the tensor coupling, $f_{\omega Y}^i$, so as to keep $\kappa_{\omega Y}$ given in the ESC08 model (see Table~\ref{tab:c.c.esc08}).  
This reduction may partly restore SU(3) symmetry and bring the present calculation close to that with SU(3) symmetry.   
We then find that the factors are as follows: $x(a) = 0.689$,  $x(b) = 0.641$,  $x(c) = 0.626$,  $x(d) = 0.683$.  Correspondingly, the $\sigma$-$Y$ coupling constant should be readjusted 
so as to reproduce the correct potential depth, $U_Y$, which is shown in column (B) in Table~\ref{tab:c.c.sY.esc08}.  
This recalculation gives the maximum neutron-star mass presented in column (B) of Table~\ref{tab:n.s.mass-su3}, namely 
the mass is reduced by about $8\%$ comparing with the result in column (A), 
and thus the mass cannot reach the value of $1.97 \pm 0.04 M_\sun$.  
Fig.~\ref{fig:pf-su3-red} shows the particle fractions, where not only the $\Xi^-$ but also the $\Lambda$ hyperon appear in the core matter. 
The threshold density for the $\Lambda$ production is $\sim 0.4 - 0.5$ fm$^{-3}$, which is 
very close to that for the $\Xi^-$.  As in cases of column (A), their threshold densities in case (a) are somewhat higher than those in cases (b) - (d).   
Therefore, the RH result again gives the heaviest neutron-star mass. 

\section{Summary and conclusions}

Using relativistic Hartree-Fock approximation, we have studied the properties of the core of neutron star in detail. In the present calculation, 
we have considered not only the tensor coupling of 
vector mesons to octet baryons and the form factors at vertexes but also the internal (quark) structure change of baryons in matter.  

We have performed the RHF calculations in two ways: one is the calculation with the coupling constants determined by SU(6) (quark model) symmetry, the other is 
with the coupling constants based on SU(3) (flavor) symmetry.  Then, from the result of the former calculation, we have found the following important features. 
The Fock term (except for the tensor coupling) makes the EoS for neutron star rather soft, while the tensor coupling pushes the threshold of hyperon 
production upward and thus the EoS becomes somewhat hard. Because the form factor reduces the effect of exchange term and the scalar self-energies for 
baryons due to the tensor coupling, 
the $\sigma$ and $\omega$ mean fields are correspondingly enhanced to keep the reasonable values of $K_v$ and the nucleon mass in matter. 
Then, the enhancement of $\omega$ mean field hardens the EoS.  Furthermore, the medium modification of the internal baryon structure also makes the EoS 
hard, and enhances the maximum neutron-star mass.   
If the $\sigma$-hyperon coupling constants are readjusted so as to reproduce the observed hyperon-potential depths (although such change eventually breaks SU(6) symmetry), 
the particle composition in the core matter is considerably modified, namely the $\Sigma$ hyperons disappear below $n_B = 1.2$ fm$^{-3}$. 
However, the calculated maximum neutron-star mass in 
SU(6) symmetry cannot reach the massive neutron-star mass, $1.97 \pm 0.04 M_\sun$ (PSR J1614-2230).

In the calculation based on SU(3) symmetry, we have used the coupling constants provided by the latest Nijmegen (ESC08) model.  
Then, it is very remarkable that the structure of particle composition in matter is completely different from that in the case of SU(6) symmetry, namely all hyperons except the $\Xi^-$ 
disappear below $n_B = 1.2$ fm$^{-3}$.  Accordingly, the EoS becomes hard, and thus, in the RH or CQMC calculation, we can obtain the mass which is consistent with 
the observed mass of $1.97 \pm 0.04 M_\sun$.  
Because, in the present calculation, the coupling constants for the non-linear $\sigma$ terms in QHD are determined so as to 
reproduce the values of $K_v$ and $M_N^*$ given by the QMC model, 
the coupling of the $\omega$ meson to the nucleon in the RH case is rather strong compared with those in the RHF cases.  
Thus, the RH calculation gives the hardest EoS, which generates the heaviest neutron-star mass, $2.0 M_\sun$.  
The medium modification of the internal baryon structure also plays the significant role in determining the neutron-star mass. 
Such effect generates the mass fraction of about $0.1 M_\sun$.  

Because the coupling constants, $g_{\sigma B}$ and $g_{\omega N}$, are determined so as to satisfy the nuclear saturation and the hyperon-potential depths at $n_0$, 
SU(3) symmetry in the coupling constants is partly broken in the present calculation.  However, such violation may be justified, because 
the ground (vacuum) state of the core matter consists of the octet baryons, which fill (positive) baryon-energy levels to the Fermi momenta, and 
each baryon fraction inside the core varies depending on the density.  It implies that SU(3) symmetry is explicitly broken in the ground state.  
However, we have checked how the neutron-star mass is modified when SU(3) symmetry is restored. 
In fact, by reducing $g_{\omega Y}$ {\em by hand}, which may (partly) restore SU(3) symmetry, we have recalculated the neutron-star mass.    
Then, we have found that the maximum mass cannot reach the desired value, $1.97 \pm 0.04 M_\sun$.  
Thus, how we determine the coupling constants, especially the coupling constants of vector mesons to baryons, is very important to obtain a heavy neutron-star mass. 

In conclusions, we have found that the extension from SU(6) symmetry to SU(3) symmetry in the meson-baryon couplings and the internal baryon-structure 
variation in matter certainly enhance the maximum neutron-star mass. Furthermore, the effects of the form factor at vertex and the Fock term including 
the tensor coupling due to the vector meson are indispensable to describe the core matter of neutron star.  
In particular, the Fock contribution is very vital to reproduce the preferable value of symmetry energy, $a_4 \, (\simeq 30 - 40$ MeV), in nuclear matter. 
Including all these effects, it becomes possible to explain the mass of PSR J1614-2230 pulsar. In the usual RH calculations with the coupling constants in SU(6) symmetry, it is 
necessary to consider the self-interaction terms of vector ($\omega, {\vec \rho}, \phi$) mesons, as well as the $\sigma$ meson, to obtain 
a heavy neutron-star mass \citep{todd,schramm,shen,bednarek,weissenborn2012a}.  
However, it should be emphasized that such self-interaction terms are {\em not} indispensable in the present RHF calculation.  

Finally, we give several comments on the future work.  
It may be important to consider the $K$ or $K^*$ meson exchange between two baryons, because they generate the mixing of 
octet baryons in matter. For example, the mixing of $N$-$\Lambda$, $\Lambda$-$\Xi$, $N$-$\Sigma$ or $\Sigma$-$\Lambda$ occurs through 
the $K$ or $K^*$ meson exchange.  Furthermore, the $\pi$ or $\rho$ meson mixes the $\Sigma$-$\Lambda$ hyperons. These mixing effects may change the baryon composition 
in the core matter.  It is also important to introduce temperature into the RHF formulation \citep{saito1989,soutome1990}. 

In the present study, we do not consider the short range (two-baryon) correlation in matter, which may affect the EoS at high density.  
In fact, using the latest ESC08 model, 
\citet{schulze} have already performed the Brueckner-Hartree-Fock calculation for the neutron-star matter, and reported very low maximum masses below 
$1.4 M_\sun$ (see also, \citet{akmal,nishizaki,logoteta}). 
Because relativity must be very crucial to investigate the high density matter, it is very desired to perform the (relativistic) Dirac-Brueckner-Hartree-Fock calculation 
including hyperons.  

At very high density, the quark and gluon degrees of freedom, rather than the hadron degrees of freedom, may take place in the core matter.  
Because the degrees of freedom in a quark-gluon matter is generally large, 
it is necessary to assume a rather strong correlation between quarks and gluons to support a massive neutron-star mass \citep{bratovic}. 
Furthermore, in the crossover between the hadron and quark-gluon phases there may exist rich non-perturbative structure such as color superconducting phases etc \citep{fukushima}.
It would be very intriguing to investigate how such degrees of freedom contribute to the EoS and the maximum mass of neutron star.

\acknowledgments

The authors (especially, T. M.) would like to thank Myung-Ki Cheoun, Tomoyuki Maruyama, T. Kajino and T. Takatsuka for fruitful discussions on the EoS for neutron stars.

\clearpage



%

\clearpage

\begin{table}
\begin{center}
\caption{\label{tab:BSE} Functions $A_{i}$, $B_{i}$, $C_{i}$, and $D_{i}$.  
The index $i$ is specified in the left column, where $V (T)$ stands for the vector (tensor) coupling at each
meson-$BB^\prime$ vertex.  The last row is for the (pseudovector) pion contribution. }
\begin{tabular}{l|c|c|c|c}
\tableline\tableline
$i$ & $A_{i}$ & $B_{i}$ & $C_{i}$ & $D_{i}$ \\
\tableline
$\sigma$ & 
$g_{\sigma B}^{2}(\bar{\sigma})\Theta_{\sigma}$ & 
$g_{\sigma B}^{2}(\bar{\sigma})\Theta_{\sigma}$ & 
$-2g_{\sigma B}^{2}(\bar{\sigma})\Phi_{\sigma}$ & 
$-$ \\
$\omega_{VV}$ & 
$2g_{\omega B}^{2}\Theta_{\omega}$ & 
$-4g_{\omega B}^{2}\Theta_{\omega}$ & 
$-4g_{\omega B}^{2}\Phi_{\omega}$ & 
$-$ \\
$\omega_{TT}$ & 
$-\left(\frac{f_{\omega B}}{2\mathcal{M}}\right)^{2}m_{\omega}^{2}\Theta_{\omega}$ & 
$-3\left(\frac{f_{\omega B}}{2\mathcal{M}}\right)^{2}m_{\omega}^{2}\Theta_{\omega}$ & 
$4\left(\frac{f_{\omega B}}{2\mathcal{M}}\right)^{2}\Psi_{\omega}$ & 
$-$ \\
$\omega_{VT}$ & 
$-$ & 
$-$ & 
$-$ & 
$12\left(\frac{f_{\omega B}g_{\omega B}}{2\mathcal{M}}\right) \Gamma_{\omega}$ \\
$\rho_{VV}$ & 
$2g_{\rho B}^{2}\Theta_{\rho}$ & 
$-4g_{\rho B}^{2}\Theta_{\rho}$ & 
$-4g_{\rho B}^{2}\Phi_{\rho}$ &
$-$ \\
$\rho_{TT}$ & 
$-\left(\frac{f_{\rho B}}{2\mathcal{M}}\right)^{2}m_{\rho}^{2}\Theta_{\rho}$ & 
$-3\left(\frac{f_{\rho B}}{2\mathcal{M}}\right)^{2}m_{\rho}^{2}\Theta_{\rho}$ & 
$4\left(\frac{f_{\rho B}}{2\mathcal{M}}\right)^{2}\Psi_{\rho}$ & 
$-$ \\
$\rho_{VT}$ & 
$-$ & 
$-$ & 
$-$ & 
$12\left(\frac{f_{\rho B}g_{\rho B}}{2\mathcal{M}}\right) \Gamma_{\rho}$ \\
$\pi_{pv}$ &
$-f_{\pi B}^{2}\Theta_{\pi}$ & 
$-f_{\pi B}^{2}\Theta_{\pi}$ & 
$2\left(\frac{f_{\pi B}}{m_{\pi}}\right)^{2}\Pi_{\pi}$ & 
$-$ \\
\tableline
\end{tabular}
\end{center}
\end{table}
\begin{table}
\begin{center}
\caption{\label{tab:parametrizationQMC} Values of $a_B$ and $b_B$ for octet baryons in the QMC or CQMC model. }
\begin{tabular}{l|cc|cc}
\tableline\tableline
\ & QMC & \ & CQMC & \ \\
\ & $a_{B}$~(fm) & $b_{B}$ & $a_{B}$~(fm) & $b_{B}$ \\
\tableline
$N$       & 0.179 & 1.00 & 0.118 & 1.04 \\
$\Lambda$ & 0.172 & 1.00 & 0.122 & 1.09 \\
$\Sigma$  & 0.177 & 1.00 & 0.184 & 1.02 \\
$\Xi$     & 0.166 & 1.00 & 0.181 & 1.15 \\
\tableline
\end{tabular}
\end{center}
\end{table}
\begin{table}
\begin{center}
\caption{\label{tab:c.c.s} Coupling constants (in SU(6) symmetry) and calculated properties of symmetric nuclear matter at $n_0$. 
The values of the incompressibility, $K_v$, the symmetry energy, $a_4$, and the self-energies for the nucleon, $\Sigma^s_N$ and $\Sigma^0_N$, are in MeV, and, 
in the parenthesis of the self-energies, the left (right) number is for the Hartree (Fock) contribution.  We note that the space component of the 
vector self-energy, $\Sigma^v_N$, is very small (at most $\sim 5$ MeV).  Note that, for case (7), the coupling constants except for $g_{\sigma Y}$ are identical to those in case (4). 
}
\begin{tabular}{c|ccccccccc}
\tableline\tableline
case &$\frac{g^2_{\sigma N}}{4\pi}$&$\frac{g^2_{\omega N}}{4\pi}$&$g_2$ (fm$^{-1}$) &$g_3$&$K_v$&$a_4$&$\frac{M_N^{\ast}}{M_N}$&$\Sigma^s_N$&$\Sigma^0_N$\\ 
\tableline
(1)&5.73&5.66&21.5&34.1&253&21.7&0.79&-193\,(-193,\,0)&-134\,(-134,\,0)\\
(2)&4.67&3.91&10.7&170&253&33.3&0.79&-193\,(-157,\,-36)&-135\,(-92,\,-43)\\
(3)&3.78&4.23&10.5&235&253&35.4&0.79&-193\,(-126,\,-67)&-133\,(-100,\,-33)\\
(4)&4.13&4.69&12.5&100&253&32.2&0.79&-193\,(-146,\,-46)&-133\,(-111,\,-22)\\
(5)&4.10&4.25&-&-&253&30.2&0.79&-193\,(-150,\,-43)&-120\,(-101,\,-19)\\
(6)&4.01&5.48&-&-&264&33.6&0.76&-226\,(-170,\,-56)&-153\,(-129,\,-24)\\
\tableline
	\end{tabular}
	\end{center}
\end{table}
\begin{table}
\begin{center}
\caption{\label{tab:n.s.mass-su6} Neutron-star radius, $R_{max}$ (in km), the central density, $n_c$ (in fm$^{-3}$), and the ratio of the maximum neutron-star mass 
to the solar mass, $M_{max}/M_\odot$ in SU(6) symmetry. 
}
\begin{tabular}{c|ccc}
\tableline\tableline
		&$R_{max}$&$n_c$&$M_{max}/M_{\odot}$\\
\tableline
		(1)&	11.5&		1.03&		1.57\\
		(2)&	10.7&		1.29&		1.36\\
		(3)&	11.4&		1.08&		1.50\\
		(4)&	11.3&		1.09&		1.52\\
		(5)&	11.2&		1.10&		1.62\\
		(6)&	11.8&		0.97&		1.67\\
		(7)&	11.9&		0.94&		1.52\\
\tableline
	\end{tabular}
	\end{center}
\end{table}
\begin{table}
\begin{center}
\caption{\label{tab:c.c.esc08} Coupling constants, $f_{\pi B}/\sqrt{4\pi}$, $g_{MB}/\sqrt{4\pi}$ and $\kappa_{MB} = f_{MB}/g_{MB}$, 
and the cutoff parameters, $\Lambda_B$, in the ESC08 model (taken from \citet{esc08}). 
}
\begin{tabular}{cc|ccccc}
\tableline\tableline
	$M$& 	&$MNN$&$M \Lambda \Lambda$&$M\Sigma \Sigma$&$M\Xi \Xi$&$\Lambda_B$(MeV){\tablenotemark{b}} \\
\tableline
	$\pi$   &	$f$ &	  0.2675 & 0      & 0.1903 & -0.0772 & 1312 \\
	$\omega$ &	$g$ &  $-{\tablenotemark{a}}$  & 2.8158 & 2.8158 & 2.0863 & 1576 \\
		& $\kappa$           & -0.2276{\tablenotemark{a}} & -1.2030 & -0.0926 & -2.1111 &  \\
	$\rho$     &	$g$ &  0.6918 & 0      & 1.3837 & 0.6918 & 969 \\
	&	$\kappa$               & 5.6805 & 0      & 2.4656 & -0.7489 &  \\
\tableline
	\end{tabular}
\tablenotetext{a}{The $\omega$-$N$ vector coupling constant is determined so as to satisfy the nuclear saturation condition (see Table~\ref{tab:prop.esc08}). 
The tensor coupling is then fixed using $\kappa_{\omega N} = -0.2276$, which is the value given in the ESC08 model. }
\tablenotetext{b}{The form factor in the ESC08 model is given by a gaussian function.  
The present values are therefore the cutoff parameters (given in the ESC08 model) multiplied by $\sqrt{2}$, 
because we use the dipole function, Eq.(\ref{formfactor}). 
We assume that the cutoff parameter at the $\sigma$-$BB$ vertex is identical to that at the $\pi$-$BB$ vertex. 
}
	\end{center}
\end{table}
\begin{table}
\begin{center}
\caption{\label{tab:prop.esc08} Coupling constants (for the ESC08 model) and calculated properties of symmetric nuclear matter at $n_0$. 
The values of $K_v$, $a_4$, $\Sigma^s_N$ and $\Sigma^0_N$ are in MeV, and, 
in the parenthesis of the self-energies, the left (right) number is for the Hartree (Fock) contribution.  Note that the space component of the 
vector self-energy is again very small as in the case of SU(6) symmetry. 
}
\begin{tabular}{c|ccccccccc}
\tableline\tableline
case &$\frac{g^2_{\sigma N}}{4\pi}$&$\frac{g^2_{\omega N}}{4\pi}$&$g_2$ (fm$^{-1}$) &$g_3$&$K_v$&$a_4$&$\frac{M_N^{\ast}}{M_N}$&$\Sigma^s_N$&$\Sigma^0_N$\\ 
\tableline
(a)&6.02&6.13&22.6&8.51&249&20.0&0.78&-204\,(-204,\,0)&-145\,(-145,\,0)\\
(b)&3.22&5.16&13.9&62.5&249&36.2&0.78&-204\,(-120,\,-85)&-144\,(-122,\,-22)\\
(c)&3.12&4.92&-&-&249&35.3&0.78&-204\,(-120,\,-84)&-135\,(-116,\,-19)\\
(d)&3.27&5.86&-&-&261&35.5&0.75&-233\,(-143,\,-90)&-163\,(-138,\,-25)\\
\tableline
	\end{tabular}
	\end{center}
\end{table}
\begin{table}
\begin{center}
\caption{\label{tab:c.c.sY.esc08} The $\sigma$-$Y$ coupling constants. For details, see the text. 
}
\begin{tabular}{c|ccc|ccc}
\tableline\tableline
		&\multicolumn{3}{c|}{(A)}&\multicolumn{3}{c}{(B)}\\
case&$g_{\sigma\Lambda}^2/4\pi$&$g_{\sigma\Sigma}^2/4\pi$&$g_{\sigma\Xi}^2/4\pi$&$g_{\sigma\Lambda}^2/4\pi$&$g_{\sigma\Sigma}^2/4\pi$&
$g_{\sigma\Xi}^2/4\pi$\\
\tableline	
		(a)&5.32&2.63&2.71&2.92&1.05&1.45\\
		(b)&7.17&3.32&3.63&3.47&1.02&1.70\\
		(c)&6.59&3.01&3.32&3.09&0.85&1.50\\
		(d)&5.22&3.04&2.50&2.77&1.14&1.29\\
\tableline
	\end{tabular}
	\end{center}
\end{table}
\begin{table}
\begin{center}
\caption{\label{tab:n.s.mass-su3} Neutron-star radius, $R_{max}$ (in km), the central density, $n_c$ (in fm$^{-3}$), and the ratio of the maximum neutron-star mass 
to the solar mass, $M_{max}/M_\odot$, in the ESC08 model.   For details, see the text. 
}
\begin{tabular}{c|ccc|ccc}
\tableline\tableline
  &\multicolumn{3}{c|}{(A)}&\multicolumn{3}{c}{(B)}\\
 case &$R_{max}$&$n_c$&$M_{max}/M_{\odot}$&$R_{max}$&$n_c$&$M_{max}/M_{\odot}$\\
\tableline
		(a)&	11.2&		1.04&		2.00&	11.6&		0.98&		1.84\\
		(b)&	11.2&		1.09&		1.83& 11.5&		1.05&		1.62\\
		(c)&	11.2&		1.06&		1.86& 11.6&		1.02&		1.71\\
		(d)&	11.5&		1.00&		1.93& 11.7&		0.99&		1.77\\
\tableline
	\end{tabular}
	\end{center}
\end{table}
%


\end{document}